\documentclass[a4paper,11pt]{article}
\usepackage{jheppub}

\usepackage[utf8]{inputenc}
\usepackage{amsmath}
\usepackage{graphicx}
\usepackage{multirow}

\newcommand{\orcid}[1]{\href{https://orcid.org/#1}{#1}}

\title{New oscillation and scattering constraints on the tau row matrix elements without assuming unitarity}

\author[1]{Peter B.~Denton\note{\orcid{0000-0002-5209-872X}}}
\author[2]{and Julia Gehrlein\note{\orcid{0000-0002-1235-0505}}}

\affiliation{High Energy Theory Group, Physics Department, Brookhaven National Laboratory, Upton, NY 11973, USA}

\emailAdd{pdenton@bnl.gov}
\emailAdd{jgehrlein@bnl.gov}

\makeatletter
\hypersetup{colorlinks=true,allcolors=[rgb]{1,0.56,0},pdftitle=\@title}
\makeatother

\abstract{
The tau neutrino is the least well measured particle in the Standard Model.
Most notably, the tau neutrino row of the lepton mixing matrix is quite poorly constrained when unitarity is not assumed.
In this paper, we identify data sets involving tau neutrinos that improve our understanding of the tau neutrino part of the mixing matrix, in particular $\nu_\tau$ appearance in atmospheric neutrinos.
We present new results on the elements of the tau row leveraging existing constraints on the electron and muon rows for the cases of unitarity violation, with and without kinematically accessible steriles.
We also show the expected sensitivity due to upcoming experiments and demonstrate that the tau neutrino row precision may be comparable to the muon neutrino row in a careful combined fit.}

\begin{document}

\maketitle

\section{Introduction}
The tau neutrino is the second to last discovered particle of the Standard Model (SM) in 2000 \cite{DONUT:2000fbd}, only followed by the Higgs boson discovery in 2012 \cite{ATLAS:2012yve,CMS:2012qbp} to complete the known particle content of the SM. Nevertheless, the tau neutrino remains one of the least studied particle of the SM even more than two decades after its discovery due to the low cross section of neutrinos and the high energies required for detection and/or production.

Under the assumption of unitarity, there is good precision in the electron neutrino row and modest precision in the other two rows of the leptonic mixing matrix, the PMNS matrix \cite{Pontecorvo:1957qd,Maki:1962mu}.
In the more general case when unitarity is not assumed, however, the tau row of the PMNS matrix, presents fairly large uncertainties \cite{Parke:2015goa,Ellis:2020ehi, Ellis:2020hus,Hu:2020oba} such that the three mass states we are familiar with may only account for $\mathcal O(80)\%$ of the tau neutrino.
The PMNS matrix is significantly less constrained than the CKM matrix \cite{Cabibbo:1963yz,Kobayashi:1973fv} for the quark sector \cite{Charles:2015gya}.
Beyond just a robustness test of the three-flavor picture, non-unitary mixing matrices arise in many extensions of the SM such as models with neutrinos propagating in extra dimensions \cite{ArkaniHamed:1998vp,ArkaniHamed:1998sj,Bhattacharya:2009nu} and in extensions of the SM with new, heavy neutrinos potentially connected to neutrino mass generation \cite{Minkowski:1977sc,Schechter:1980gr,Foot:1988aq}. These scenarios lead to \emph{apparent} low energy unitarity violation (UV) as the mixing matrix of the full theory, including the often kinematically inaccessible sterile neutrinos, is unitary.

In the following we will not consider the case where the mixing matrix of the full theory is not unitary, rather we focus on two scenarios: UV coming from kinematically accessible sterile neutrinos whose mass is large such that their oscillations are averaged out at the detectors of neutrino oscillation experiments, and the case of kinematically inaccessible sterile neutrinos whose mass is too large to be produced in the neutrino source. While in the first case UV constraints provide complementary (and more general) bounds on the sterile neutrino parameters which can be compared to constraints from direct searches for steriles, kinematically inaccessible sterile neutrinos at oscillation experiments can only be directly constrained by high energy collider searches for sterile neutrinos with masses up to $\sim 100 $ GeV (for a recent review of sterile neutrino bounds across many energy scales see \cite{Bolton:2019pcu}).
In fact, UV is the only detectable signature of heavy sterile neutrinos with mass $m_4\gtrsim 1$ TeV, beyond which direct searches lose sensitivity. Since these experiments rely on the detection of the decay products of sterile neutrinos, these experiments are generally insensitive to light steriles with mass $m_4\lesssim 10$ MeV which corresponds to decay length outside of the detector\footnote{This is true for the coupling of steriles to $\nu_\mu$ and $\nu_\tau$. The sterile coupling to $\nu_e$ can be constrained by peak searches in nuclear decays which allow to probe smaller sterile masses down to the eV scale \cite{Bolton:2019pcu}.}.
Therefore UV constraints from oscillations can also be converted into bounds in unexplored regions of sterile neutrino parameter space.
Hence probing UV constrains a vast parameter space and is a robust way of testing for sterile neutrinos and heavy neutral leptons across many energy scales.

While the constraints on UV in the electron row are already rather tight and will continue to improve
\cite{Ellis:2020ehi,Ellis:2020hus,Hu:2020oba,Qian:2013ora,Forero:2021azc} due to the very precise measurements of reactor experiments, the muon and tau rows currently allow for sizable $\mathcal O(10\%)$ deviations from unitarity. The muon row is mostly informed from muon disappearance and electron appearance at long baseline experiments like NOvA and T2K as well as atmospheric neutrinos measured by Super-Kamiokande and IceCube. On the other hand so far only tau appearance data in the long baseline experiment OPERA and from atmospheric muon disappearance has been considered to constrain the tau row. Together with the conditions from the unitarity of the full matrix constraints on the tau row has been derived in the literature for the two scenarios we will consider in the following. However due to the small size of the tau neutrino data set the constraints are rather poor. 

In this paper we will identify new tau neutrino oscillation data sets\footnote{We will focus on oscillation and scattering experiments here but it is also possible to constrain UV using electroweak precision data \cite{Antusch:2006vwa,Fernandez-Martinez:2015hxa,Fernandez-Martinez:2016lgt}.} and show that they provide powerful constraints of the individual elements of the tau row in a UV framework.
We find, in fact, the tau data sets are richer than previously assumed which already leads to a noticeable improvement of the constraints with current data and improvements in the future. Specifically we take results from atmospheric tau neutrino appearance, astrophysical tau neutrino appearance, and charged current scattering experiments into account.
Additionally, we investigate the impact of neutral current measurements to constrain UV subject to theoretical predictions. Even though in these experiments the flavor of the neutrino is not identified, compelling bounds arise from a reduction of the total neutrino flux and non-trivial cross section modifications due to UV which can be compared to theoretical predictions.

After introducing the UV framework to calculate the oscillation probabilities including the matter effect in section \ref{sec:UV_framework}, we will show how these data sets compare in their ability to constrain the tau row when unitarity is not assumed for two benchmark cases: with and without kinematically accessible steriles.
We also forecast future sensitivities in section \ref{sec:results} and show that with the advent of new experiments and more data improvements are anticipated such that the tau row can be potentially constrained comparably as well as the other rows are now.
As we want to focus on the tau row only and establish the use of these new data sets, we will not conduct a full global fit including the electron and muon row but use priors on these rows from the literature. 
We discuss our results and conclude in sections \ref{sec:discussion} and \ref{sec:conclusions}.
Appendix \ref{sec:exp} contains more information about the experiments considered in the analysis and their implementation.

\section{Unitary violation framework}
\label{sec:UV_framework}
A UV framework is a relatively model independent framework to quantify how much the $3\times3$ lepton mixing matrix deviates from unitary.
Apparent unitary violation can manifest itself out of a number of complete models, often related to neutrino mass generation.
For concreteness, we focus on models that are parameterized as additional singlet fermions that may or may not be kinematically accessible in the typical decays that produce neutrinos, but do not lead to new frequencies ($\Delta m^2$'s) that are directly accessible through oscillations.

In this section we first present our framework and define and justify our focus on two particular UV schemes.
We then move on to calculating the physical observables in each scheme, in particular the oscillation probabilities including the matter effect, as well as cross sections.

\subsection{Unitary violation schemes}
We parameterize UV in terms of a larger $n\times n$ unitary matrix for which $m$ states are kinematically accessible in a given production channel, see table \ref{tab:accessible} for a summary of the kinematics of different relevant neutrino sources.
We then describe a given scenario by the pair of numbers: $(n,m)$, that is: (total number of neutrinos, number of accessible neutrinos).
So the standard oscillation picture is (3,3).
The case with one extra light ($m_4\sim10$ eV) sterile neutrino that is oscillation averaged in all relevant experiments is (4,4).
The case with three additional heavy ($m_4\gtrsim50$ MeV) states would be (6,3).
We focus on two cases of phenomenological interest: (4,4) and (5,3), although there are numerous others that are distinct from these.

The (4,4) case is the one with one new mass eigenstate that is accessible.
This is well motivated phenomenologically due to a large number of searches for this scenario in the averaged out limit along with some very slight hints for this case \cite{Mention:2011rk,Giunti:2010zu,Kostensalo:2019vmv}\footnote{This case should not be confused with the ``light sterile'' scenario which features kinematically accessible sterile neutrinos ($\Delta m^2\sim 1$ eV) whose oscillations can be resolved at experiments, also in this case some slight hints exist \cite{LSND:2001aii,MiniBooNE:2018esg,IceCube:2020tka}.}.
The (4,4) case corresponds to a $4\times4$ unitary matrix which, after rephasing, can be parameterized with 9 free parameters: 6 angles and 3 phases\footnote{A general $4\times4$ unitary matrix has 16 real parameters, 7 of which are removed by rephasing of the charged and neutral leptons in the ultrarelativistic limit.}.

The (5,3) case is the one with two new heavy mass eigenstates that are kinematically inaccessible. This case is top-down motivated when heavy sterile neutrinos (also known as heavy neutral leptons) are involved in the light neutrino mass generation which requires at least two generations of steriles to be in agreement with the experimental data on light neutrino masses.
This case is further motivated in leptogenesis models which, in the simplest realization of a high-scale leptogenesis mechanism \cite{Fukugita:1986hr}, require at least two generations of sterile neutrinos with masses above the weak scale.
We focus on (5,3) as opposed to (4,3) or (6,3) since the extra degrees of freedom in (6,3) are not accessible in standard oscillation experiments.
That is, the (5,3) case has enough degrees of freedom such that all elements of the directly probable $3\times3$ matrix can vary freely, unlike in the (4,3) case. Hence the (5,3) case corresponds to a non-unitary (3,3) matrix which is a submatrix of the complete, unitary mixing matrix. This case can be parameterized with 13 real parameters corresponding to 9 angles and 4 phases, see e.g.~\cite{Blennow:2016jkn}.

We parametrize these two scenarios as
\begin{align}
U=U_{35}U_{25}U_{15}U_{34}U_{24}U_{14}U_{23}U_{13}U_{12}\,,
\label{eq:Us}
\end{align}
where all matrices involving sterile states ($i=4,5$) involve a complex phase and an angle whereas the SM phase $\delta$ is contained in $U_{13}$; $U_{12}$ and $U_{23}$ are real matrices.
For (4,4) all matrices in eq.~\ref{eq:Us} involving the 5th mass state are unity.
With this parametrization we return to the standard three flavor mixing matrix if all sterile angles are zero.

\begin{figure}
\centering
\includegraphics[width=0.48\textwidth]{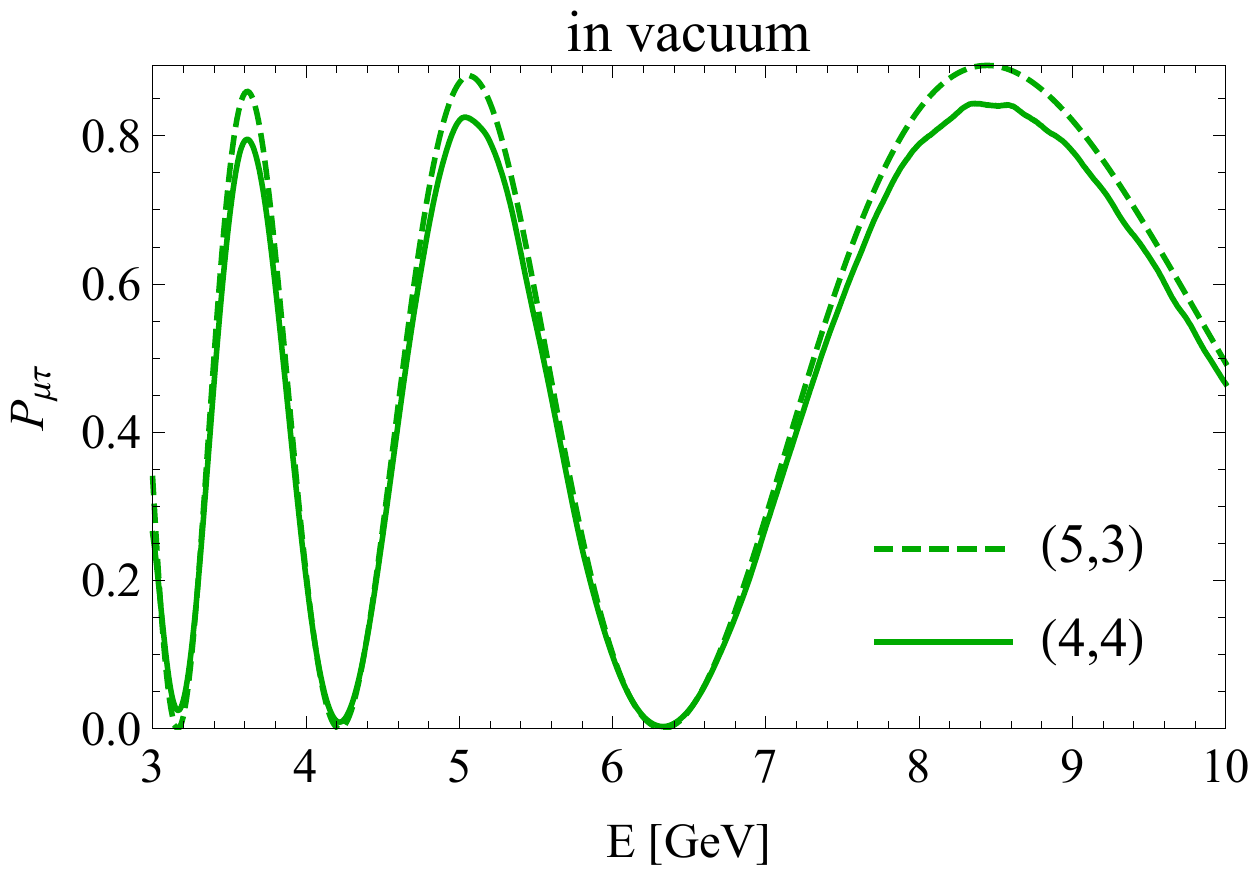}
\includegraphics[width=0.48\textwidth]{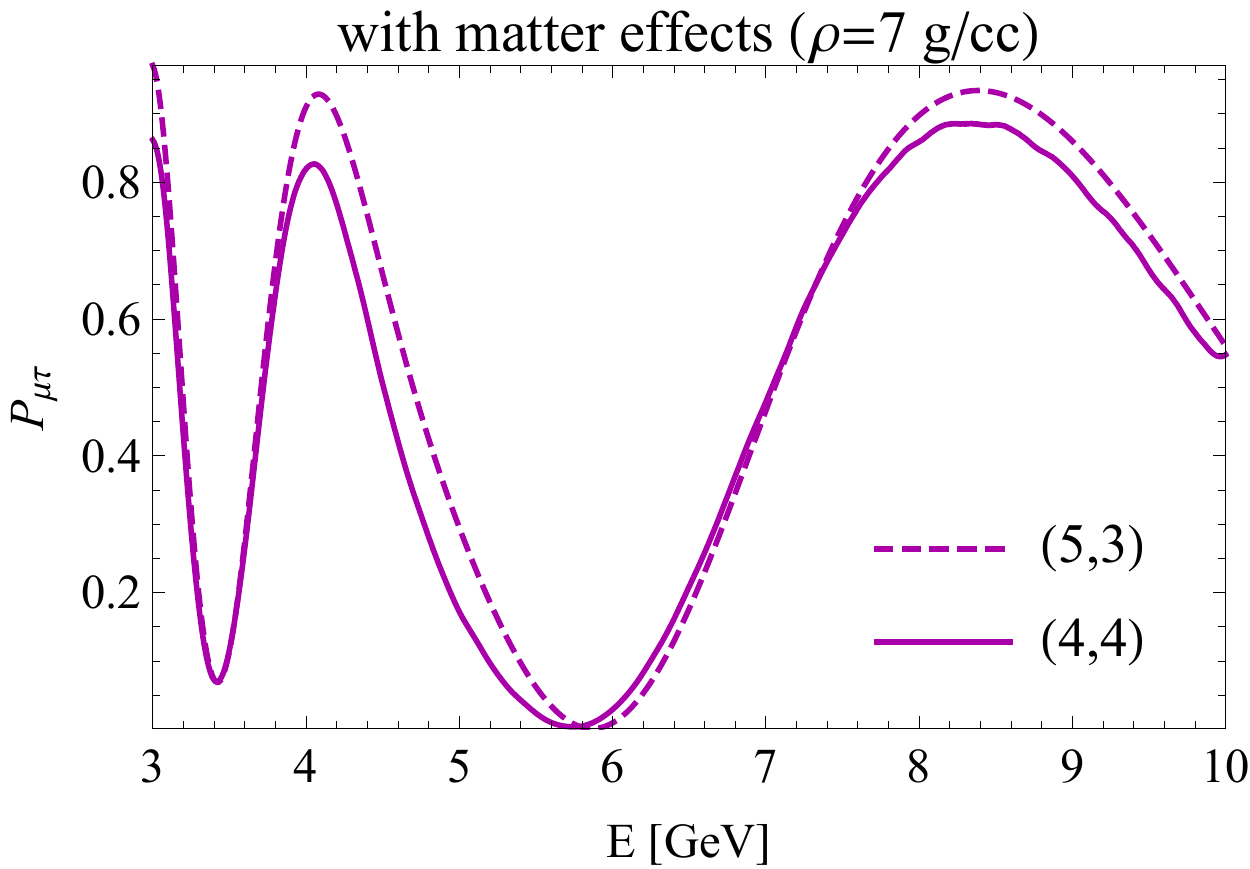}
\caption{Comparison of the tau neutrino appearance probability $P_{\mu\tau}$ in the (4,4) and (5,3) scenario with all SM angles at their best fit values from \cite{Esteban:2020cvm}, the sterile angles are $\theta_{i4}=10^\circ,~\theta_{i5}=0$ and all phases are 0. The left plot shows the probabilities for a baseline of the Earth's diameter in vacuum, the right plot takes matter effects with $\rho=7$ g/cc into account.
The (4,4) curves have the fast frequencies smoothed out.}
\label{fig:probcomp}
\end{figure}

As it has been shown in \cite{Fong:2016yyh,Blennow:2016jkn,Fong:2017gke}
for small mixing angles $\theta_s$ the oscillation phenomenology of these two benchmark scenarios is exact up to $(\theta_s)^4$. Therefore one would conclude that for a single experiment in vacuum, (4,4) and (5,3) would behave quite similarly.
Several additional differences exist in reality, however.
One simple one is that with (5,3) there are more parameters in the fit than with (4,4).
But even at the oscillation probability level there is an additional difference due to the matter effect which manifests differently in each case.
In fig.~\ref{fig:probcomp} we show one such example where in vacuum the primary difference between (4,4) and (5,3) is the amplitude of the oscillations which could be mitigated by changing the oscillation parameters or adjusting a flux normalization.
In matter, however, this does not hold as the location of the oscillation maxima and minima also change as well as the overall shape as shown in fig.~\ref{fig:probcomp}.
We have verified that this trend holds generally and, depending on the various complex phases, can have even more pronounced shape differences in addition to changes in the amplitude and location of the extrema.

While for the $n\gneq m$ scenarios, the heavy neutrinos cannot be directly produced, in principle, with a large number of extremely precise measurements of $\nu_e$, $\nu_\mu$, and $\nu_\tau$ oscillations with the known $\Delta m^2$'s, it is possible to differentiate between (3,3), (4,3), and (5,3), but (5,3) cannot be differentiated from (6,3) or scenarios with more heavy sterile neutrinos.
To see this, we note that there are 9 SM oscillation channels: 6 appearance channels and 3 disappearance channels.
Of the 6 appearance channels 3 channels are related via T or CP such that they constrain only the complex phases. This means there are 6 channels to constrain the absolute values of the matrix elements. That is enough to constrain the (3,3) and (4,4) cases. However in the (5,3) case we have more free parameters. This means that we cannot directly measure all matrix elements however we can make consistency check to see if the data fits the (5,3) case or the (4,4) case (or (4,3)) hence providing a distinction between these cases and motivates our study of them as benchmark cases.

Table \ref{tab:accessible} shows the approximate kinematic range for the new mass states considered for each scenario: (4,4) and (5,3) in all relevant experiments.
For the lower limit in the (4,4) scenario we require averaged out oscillations with $\Delta m_{41}^2 L/E\gg 1$. For the upper limit we require that the width of a decaying particle into a sterile neutrino is not more than $\sim10\%$ smaller than the decay rate into an active neutrino. This assures that the neutrino production spectrum is not significantly affected by the presence of a sterile neutrino and we can use the same approach in calculating the expected number of events in all scenarios under consideration.
If the neutrinos are produced via pion decays this bound is 15 MeV, if they are produced via $D_s$ meson decays this bound is 90 MeV.
For solar neutrinos the upper bound is 5 MeV from the maximal energy of the $^8$B flux which is 20 MeV.
For astrophysical or solar neutrinos there is no lower bound as their baselines are so large that they are oscillation averaged\footnote{We assume that the new mass states are heavier than the known states.
That is, that we are in the 3+1 or 3+2 hierarchies and not a 1+3 type hierarchy which is disfavored from the cosmological sum of neutrino masses \cite{Planck:2018vyg}.}.
For the (5,3) scenario we obtain only a lower bound on the sterile mass above which the steriles are kinematically inaccessible. This means they cannot be produced in decays of pions or $D_s$ mesons. For pions this mass is $m_4\gtrsim m_\pi-m_\mu\approx 40$ MeV, for $D_s$ mesons $m_4\gtrsim m_{D_s}-m_\tau\approx 200$ MeV.
Additional sources of tau neutrinos include those directly from $\tau$ lepton decays which leads to a higher available phase space $\sim1.5-1.8$ GeV \cite{Ballett:2019bgd}, but is not a dominant channel for the experiments we are considering.

\begin{table}
\centering
\caption{Overview of the experiments considered in the following and the relevant mass scales for the heavy states in the two UV scenarios considered.}
\begin{tabular}{c||c|c}
experiment&(4,4) ($ m_{4}$)&(5,3) ($ m_{4}$)\\\hline
atmospheric $\nu_\mu$ disappearance&$\in[10~\text{eV},15$ MeV]&$\gtrsim 40$ MeV\\
atmospheric $\nu_\tau$ appearance &$\in[10~\text{eV} , 15$ MeV]&$\gtrsim 40$ MeV\\
astrophysical $\nu_\tau$ appearance& $\lesssim 15$ MeV&$\gtrsim 40$ MeV \\
solar data&$\lesssim$ 5 MeV &$\gtrsim20$ MeV \\
DONuT/FASERnu&$\in[100~\text{eV}$, 90 MeV]&$\gtrsim 200$ MeV\\
LBL $\nu_\tau$ appearance (OPERA)&$\in[1~\text{eV}$, 15 MeV]&$\gtrsim 40$ MeV\\
LBL $\nu_\tau$ appearance (DUNE)& $\in[ 0.1$ eV, 15 MeV]&$\gtrsim40$ MeV\\
LBL $\nu_\mu$ disappearance (DUNE)& $\in[ 0.1$ eV, 15 MeV]&$\gtrsim 40$ MeV\\
CEvNS& $\in[10$ eV, 15 MeV]&$\gtrsim 40$ MeV
\end{tabular}
\label{tab:accessible}
\end{table}

\subsection{Unitary violation observables}
We now review the vacuum oscillation probability in the absence of unitarity which we will utilize in the following section for our numerical results.
The oscillation probability from flavor $\alpha$ to flavor $\beta$ with a non-unitary mixing matrix $U$ is \cite{Antusch:2006vwa} 
\begin{align}
 P_{\alpha\beta}(E,L)=\frac{|\sum_{i=1}^{acc}U_{\alpha i}^*\text{e}^{\text{i}P_iL}U_{\beta i}|^2}{(U U^\dagger)_{\alpha\alpha}(U U^\dagger)_{\beta\beta}}\,,
 \label{eq:osciprob}
\end{align}
where the sum in the numerator is over all kinematically accessible mass eigenstates $i$ with momentum eigenvalues $P_i=\sqrt{E_i^2-m_i^2}$.
The terms in the denominator $(U U^\dagger)_{\alpha\alpha}$ should be understood as $(U U^\dagger)_{\alpha\alpha}=\sum_{i=1}^{acc}U_{\alpha i} U^*_{i\alpha}$ where the sum is over kinematically accessible states. If all mass states are kinematically accessible this term sums to 1 and thus this term does not play a role in the (4,4) case, however it is crucial in the (5,3) case and drives the constraints.

The charged current (CC) cross sections and fluxes also get modified due to the non-unitary mixing matrix such that the measured cross section and flux are functions of the expected cross section in the SM ($\sigma^{CC,SM},~\phi^{CC,SM}$),
\begin{align}
 \sigma^{CC}_\alpha=\sigma^{CC,SM}_\alpha (UU^\dagger)_{\alpha\alpha}\,,\qquad
 \frac{d\phi^{CC}_\alpha}{dE}=\frac{d\phi^{CC,SM}_\alpha}{dE} (UU^\dagger)_{\alpha\alpha}~.
 \label{eq:ccxsec}
\end{align}
The neutral current (NC) cross sections get modified as
\begin{align}
 \sigma^{NC}_i=\sigma^{NC,SM}\sum_{j=1}^{acc} |(U^\dagger U)_{ij}|^2\,,
\end{align}
where $(U^\dagger U)_{ij}=\sum_{\alpha=e}^{\tau}U^*_{\alpha i} U_{\alpha j}$ summed over all active flavors.
in the flavor basis
\begin{align}
\sigma^{NC}_\beta=\sigma^{NC,SM} |(UU^\dagger )_{\beta\beta}|^2\,,
\end{align}
summed over all accessible mass states as before.
All together the number of measured events of flavor $\beta$ coming from a beam of neutrinos with flavor $\alpha$ is 
\begin{align}
 n^{meas}_\beta\sim\int dE \frac{d\phi_\alpha(E)}{dE}P_{\alpha\beta}(E,L)\sigma_\beta(E)\epsilon(E)\,,
\end{align}
with the detection efficiency $\epsilon$.
In many experiments the production and detection are both CC processes so the number of events is
\begin{align}
 n^{CC}_\beta\sim\int dE \frac{d\phi_\alpha^{CC,SM}(E)}{dE}\tilde{P}_{\alpha\beta}(E,L)\sigma_\beta^{CC,SM}(E)\epsilon(E)\,,
 \label{eq:ncc}
\end{align}
where we define $\tilde{ P}_{\alpha\beta}$
\begin{align}
 \tilde{P}_{\alpha\beta}=|\sum_{i=1}^{acc}U_{\alpha i}^*\text{e}^{\text{i}P_iL}U_{\beta i}|^2\,,
 \label{eq:osciprob_tilde}
\end{align}
such that the extra terms from eq.~\eqref{eq:ccxsec} cancel the denominator in the probability from eq.~\eqref{eq:osciprob}.
In this case the only impact of the UV matrix elements is in $\tilde{P}_{\alpha\beta}$.
On the other hand for NC detection processes the denominators do not cancel.
Then the number of detected events of any flavor assuming the probability is to be written in the form $\tilde P_{\alpha i}$ as for solar neutrinos due to the adiabatic flavor changing MSW effect \cite{Mikheyev:1985zog}, is
\begin{align}
 n^{NC}_{solar}\sim\int dE \sigma^{NC,SM}\sum_{\alpha=e}^{\tau} \frac{d\phi_\alpha^{CC,SM}(E)}{dE} \sum_{i=1}^{acc}\tilde{P}_{\alpha i}(E,L)\sum_{j=1}^{acc} |(U^\dagger U)_{ij}|^2\epsilon(E)\,,
 \label{eq:nnc solar}
\end{align}
while for coherent elastic neutrino nucleus scattering (CEvNS) it is
\begin{align}
n^{NC}_{CEvNS}\sim\int dE\sigma^{NC,SM}\sum_{\alpha=e}^\tau\frac{d\phi_\alpha^{CC,SM}(E)}{dE}\sum_{\beta=e}^\tau\tilde P_{\alpha\beta}(UU^\dagger)_{\beta\beta}\epsilon(E)\,,
\label{eq:nnc cevns}
\end{align}
where we note the presence of one extra factor of $(UU^\dagger)_{\beta\beta}$ due to the fact that this is a NC detection and there is an outgoing neutrino.

Hence in NC processes the UV matrix elements appear in several places, making these experiments a different probe of UV.
We present in appendix \ref{sec:sum_exp} a summary of the experiments and determine if there is a cancellation of the terms in the denominator or not in the different data sets.

Additionally, in the presence of UV in the (5,3) case the Fermi constant obtains corrections as not all mass states can get produced, the relation between $G_F^U$, extracted from muon decay, and $G_F^{UV}$ is
\begin{align}
 G_F^{UV}=\frac{G_F^U}{\sqrt{(UU^\dagger)_{ee}(UU^\dagger)_{\mu\mu}}}\,.
 \label{eq:gf}
\end{align}

Even though the change in the Fermi constant depends only on elements in the electron and muon row when deriving constraints on the tau row the muon row elements are involved such that we often predict a deviation from $G_F^U$ for our constraints on the tau row.
The matter effect must also be carefully included in the oscillations, see \cite{Denton:2021rsa,Fernandez-Martinez:2007iaa,Fong:2017gke}.
In fact, as we will show in the following the matter effect plays an important role in constraining the tau row, in particular for channels involving atmospheric neutrinos like atmospheric muon disappearance and atmospheric tau appearance.

\section{Results}
\label{sec:results}
Here we will present the numerical results for the (4,4) and (5,3) case. 
We present tau data sets which can be grouped in different categories, for more details see appendix \ref{sec:exp}.
\begin{itemize}
\item \textbf{Atmospheric muon disappearance experiments}: These experiments are sensitive to the mixing of heavy sterile neutrinos with tau neutrinos due to the presence of the SM matter effect \cite{Blennow:2018hto}.
 We use current constraints from DeepCore \cite{Aartsen:2017bap} and Super-Kamiokande \cite{Abe:2014gda} on the mixing between tau neutrinos and a sterile neutrino.
This constraint will improve in the future by the successors of these experiments as well as KM3NeT/ORCA \cite{Aiello:2021aqp}\footnote{In addition to atmospheric muon disappearance experiments also long baseline experiments can constrain the tau row with muon disappearance and NC measurements \cite{MINOS:2017cae,NOvA:2017geg,NOvA:2021smv,T2K:2019efw,Forero:2021azc} however these constraints are slightly weaker.}.
\item \textbf{Atmospheric tau appearance experiments}: The sensitivity to atmospheric tau appearance comes primarily from a combination of the matter effect, the lower tau neutrino reconstructed energy, and the rising cross section due to the tau lepton's threshold \cite{Denton:2021rsa}.
IceCube \cite{Aartsen:2019tjl} and Super-Kamiokande \cite{Li:2017dbe} have constrained the tau normalization $N_\tau$ defined as the ratio of the measured $\nu_\tau$ flux to the expected one assuming standard oscillations. This constraint will improve in the future with IceCube-Gen2 \cite{Ishihara:2019aao}, KM3NeT/ORCA \cite{Eberl:2017plv,Aiello:2021jfn} and Hyper-Kamiokande \cite{Abe:2018uyc} data.
\item \textbf{Astrophysical tau appearance experiments}: Astrophysical neutrino sources only produce electron or muon neutrinos\footnote{The matter effect in the source could induce oscillations, but only for energies $\lesssim100$ GeV \cite{Razzaque:2009kq}, well below the region of interest for IceCube. In addition, a subleading component of intrinsic $\nu_\tau$ could be present depending on the source, but will not contribute a considerable fraction of the total flux.}, hence the observation of astrophysical tau neutrinos indicates a flavor change \cite{Palladino:2018qgi}. From the ratio of currently observed astrophysical tau neutrinos to astrophysical muon neutrinos at IceCube \cite{Abbasi:2020zmr, Stettner:2019tok} we obtain constraints on the tau row. IceCube-Gen2 will measure this ratio even more precisely \cite{IceCube:2014gqr}.
\item \textbf{Tau neutrino appearance at long baseline experiments}: OPERA \cite{OPERA:2018nar} as well as DUNE in the future are sensitive to tau appearance in a muon neutrino beam \cite{deGouvea:2019ozk} which provides insights on the tau matrix elements. 
\item \textbf{Charged current and neutral current scattering experiments}:
Charged current scattering experiments such as NOMAD/CHORUS \cite{Astier:2001yj, Eskut:2000de}\footnote{Notice that NOMAD/CHORUS do not lead to constraints on the tau row matrix elements unless unitarity is violated in the electron or muon row, see appendix.} as well as DONuT \cite{Kodama:2007aa} and, in the future, FASERnu \cite{Abreu:2019yak} and other future forward physics facilities \cite{Anchordoqui:2021ghd} identify tau neutrinos at very short baselines where no SM oscillations have developed yet. On the other hand, neutral current scattering experiments like coherent elastic neutrino nucleus scattering (CEvNS) experiments or SNO do not identify the tau neutrinos but they are sensitive to all neutrino flavors. As demonstrated in the previous section and in eqs.~\eqref{eq:nnc solar} and \eqref{eq:nnc cevns} these observations still provide some constraints on the tau row.
\end{itemize}

As we are focused on the tau row and demonstrate that several new data sets lead to an improved knowledge of the tau matrix elements, we use priors on the matrix elements of the electron and muon row instead of conducting a full global fit.
For the (4,4) case we use the current constraints on the electron and muon row from the recent global fit in \cite{Hu:2020oba}, for the (5,3) case we use the constraints from the recent global fit in \cite{ Ellis:2020hus,Ellis:2020ehi} whose results have been explicitly derived in the benchmark scenarios we study. These publications also included tau data from OPERA, NOMAD, and atmospheric muon disappearance in their global fit. These data sets also affect the muon row\footnote{The electron row is also affected by these tau data however mostly in an indirect way via the unitarity triangles. Additionally the constraints on the electron row are mainly driven by reactor experiments which are not sensitive to tau neutrinos.}, so the worry of double counting might arise. However the strongest constraints on the muon row come from muon neutrino disappearance experiments like the long-baseline experiments NOvA and T2K as it can be seen from the small level of improvement of the muon row between older publications in the literature before appearance data existed such as \cite{Parke:2015goa}, to the more recent publications \cite{Ellis:2020ehi, Ellis:2020hus,Hu:2020oba}. Furthermore, the currently used tau data sets have lower statistics than the muon data sets such that they present larger uncertainties. 
Nevertheless, we caution the reader that a direct comparison to global analyses is potentially subject to small corrections due to double counting.
We present our results in the context of the constraints derived from information in the $\nu_e$ and $\nu_\mu$ rows only, the addition of each class of experiments, and finally the sum of all constraints to highlight the impact on the tau neutrino row for each class of experiment.

For the forecasted results we use the same priors on the electron and muon rows. This can be understood as a conservative approach and will showcase the improvements from the new tau data sets alone. 
It should be noted though that the electron and muon rows will improve as well with future data from the JUNO, DUNE, HK, and IceCube-Gen2 experiments \cite{Qian:2013ora, Ellis:2020ehi,Ellis:2020hus}.

Our parameterization of the mixing matrix in both scenarios involves phases in the mixing matrix. However the sensitivity of the new tau data sets to CP violation is rather low and as in general there is no strong preference for any value of the phase in the (3,3) scenario \cite{Esteban:2020cvm} we will constrain all phases to be CP conserving in our analysis\footnote{In fact, there is currently a slight disagreement for the preferred value of the CP violating phase in the (3,3) scenario between NOvA \cite{NOvA:2021nfi} and T2K \cite{T2K:2021xwb} which can be resolved with the introduction of new non-standard matter effects \cite{Denton:2020uda,Chatterjee:2020kkm}.}. This assumption is also consistent with the findings of \cite{Ellis:2020ehi, Ellis:2020hus} which prefer CP conserving phases for the (5,3) case. This also allows us to treat neutrinos and anti-neutrinos in the same way in our analysis.
Future analyses will need to account for all possible complex phases as NOvA and T2K data improve and DUNE and HK come online.

\begin{figure}
\centering
\includegraphics[width=\textwidth]{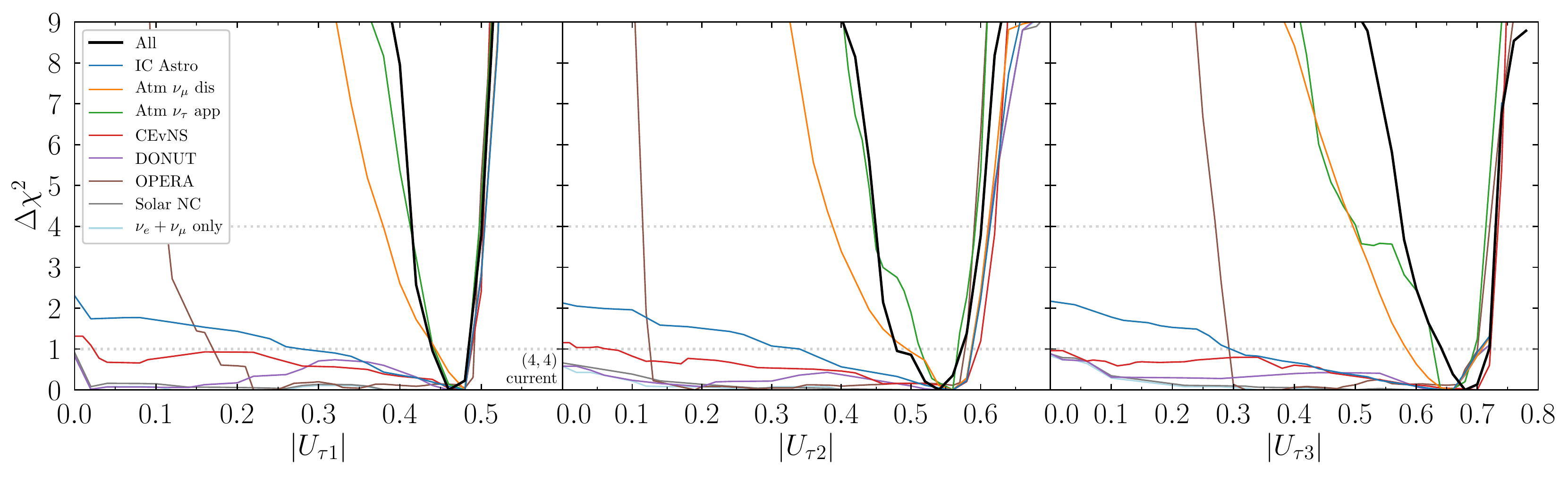}
\caption{Constraint on the individual matrix elements in the tau row in the (4,4) case using the currently available data as well as priors on the electron and muon row elements from \cite{Hu:2020oba}. The different colors represent different data sets which have been included in addition to the constraints from unitarity using priors from the electron and muon row. The black lines represents the constraints using all data sets.}
\label{fig:uti44_current}
\end{figure}

\begin{figure}
\centering
\includegraphics[width=\textwidth]{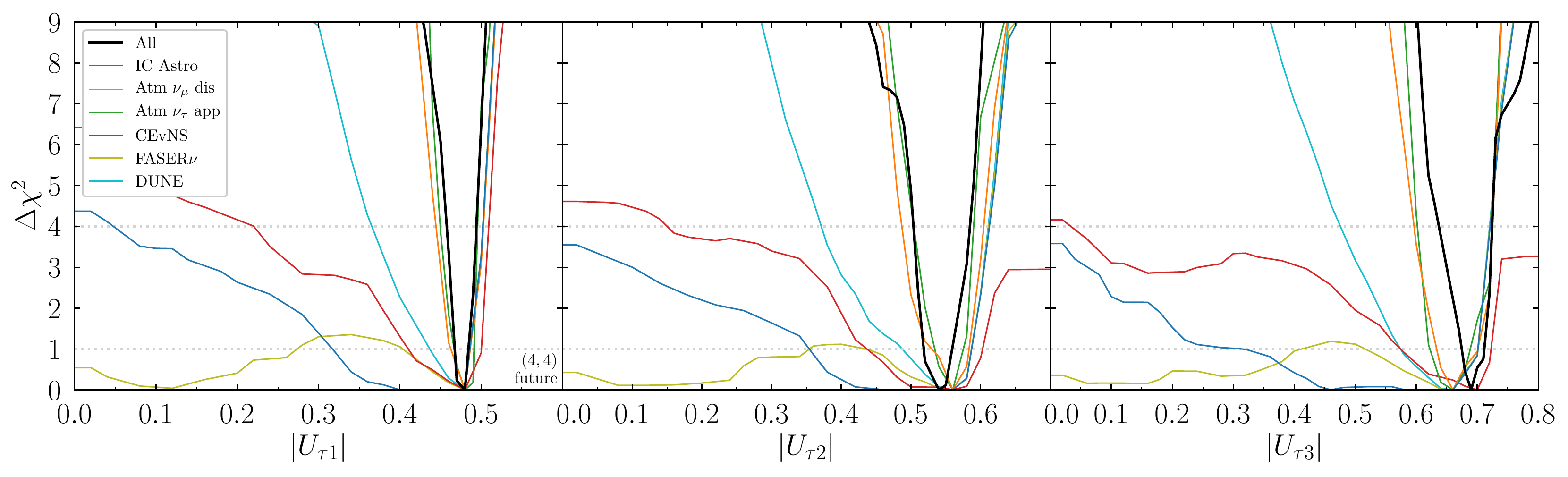}
\caption{Constraint on the individual matrix elements in the tau row in the (4,4) case using forecasted data as well as priors on the electron and muon row elements from the current constraints in \cite{Hu:2020oba}. The different colors represent different data sets which have been included in addition to the constraints from unitarity using priors from the electron and muon row. The black lines represents the constraints using all data sets.}
\label{fig:uti44_future}
\end{figure}

Finally, we use a simple $\chi^2$ test statistic depending on the expected number of events $n^{theo}_{events}$ under the UV hypothesis and the measured number of events $n^{meas}_{events}$ with uncertainty $\sigma_{meas}$
\begin{align}
 \chi^2=\frac{(n^{theo}_{events}-n^{meas}_{events})^2}{\sigma_{meas}^2}~.
\end{align}
For some experiments we also include nuisance parameters which we then minimize the test statistic over, see appendix \ref{sec:exp} for more experiment specific details.
We do not take any correlations between different experiments or theory predictions into account.

\begin{figure}
\centering
\includegraphics[width=\textwidth]{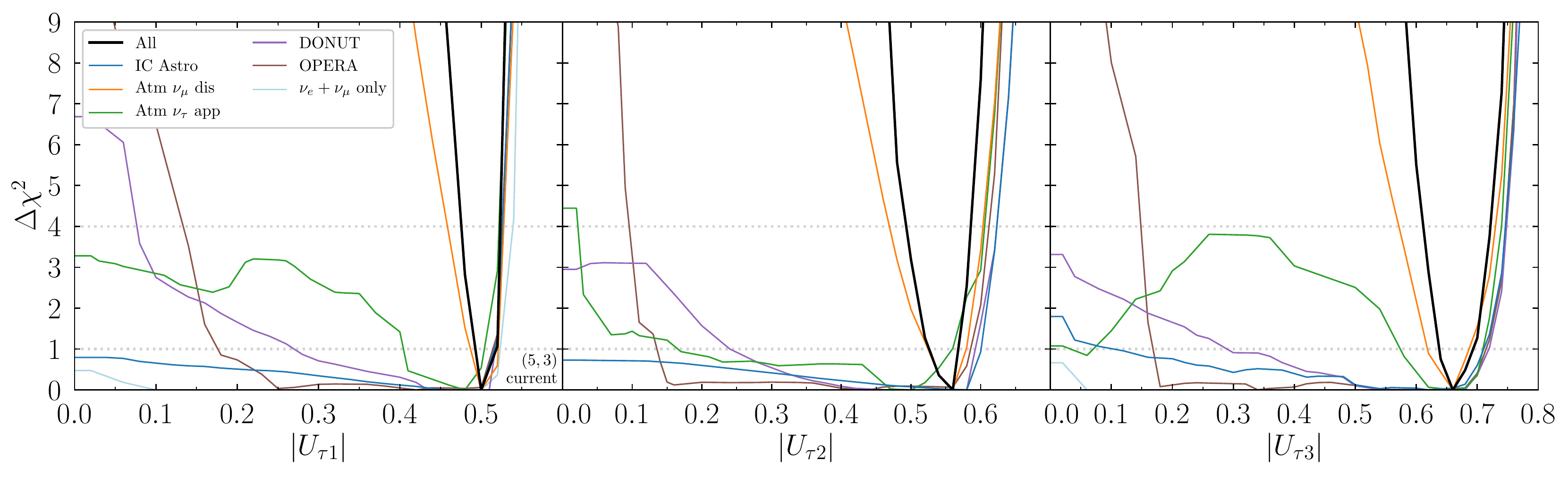}
\caption{Constraint on the individual matrix elements in the tau row in the (5,3) case using the currently available data as well as priors on the electron and muon row elements from \cite{Ellis:2020ehi,Ellis:2020hus}. The different colors represent different data sets which have been included in addition to the constraints from unitarity using priors from the electron and muon row. The constraints from solar NC and CEvNS have been omitted as they are provide only mild constraints with $\Delta \chi^2\lesssim 1$.
The black lines represents the constraints using all data sets.}
\label{fig:uti53_current}
\end{figure}
\begin{figure}
\centering
\includegraphics[width=\textwidth]{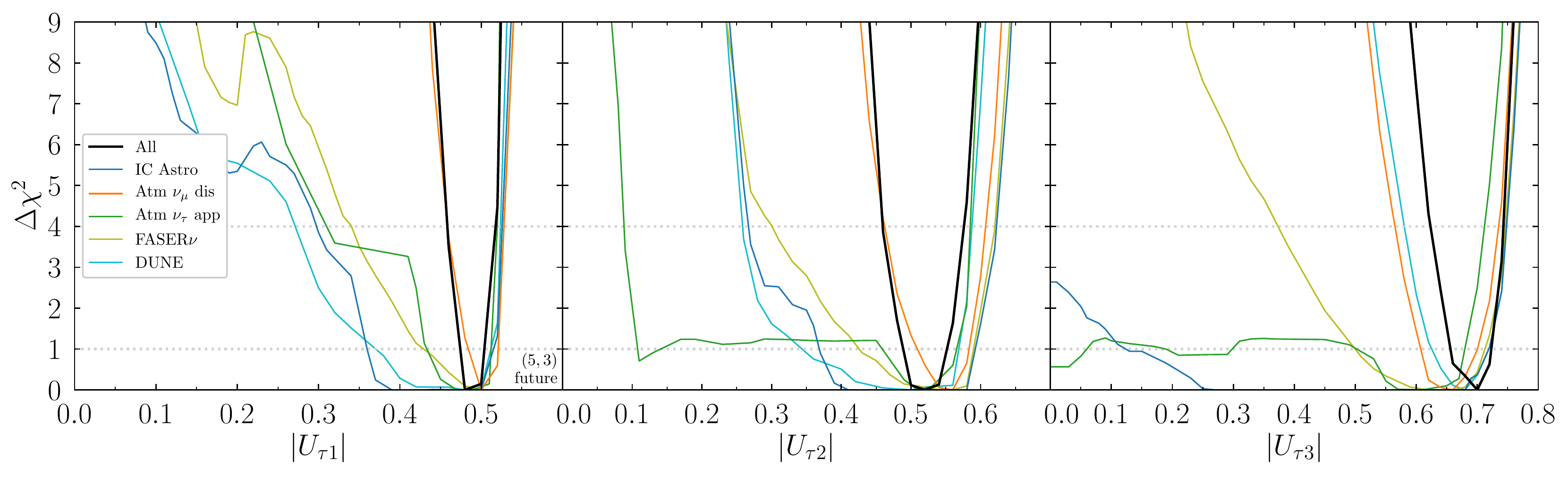}
\caption{Constraint on the individual matrix elements in the tau row in the (5,3) case using forecasted data as well as priors on the electron and muon row elements from the current constraints in \cite{Ellis:2020ehi,Ellis:2020hus}. The different colors represent different data sets which have been included in addition to the constraints from unitarity using priors from the electron and muon row.
The constraints from CEvNS have been omitted as they are provide only mild constraints with $\Delta \chi^2\lesssim 1$.
The black lines represents the constraints using all data sets.}
\label{fig:uti53_future}
\end{figure}

We present our results for the tau row matrix elements in the (4,4) case with current and forecasted in figs.~\ref{fig:uti44_current}, \ref{fig:uti44_future}, and for the (5,3) case in figs.~\ref{fig:uti53_current}, \ref{fig:uti53_future}. In each figure we show the constraints using unitarity constraints from the electron and muon row only (i.e.~no dedicated tau data has been introduced), for unitarity plus one tau data set, and for all tau data sets together. Our results demonstrate that the tau matrix elements are non-zero at a high level of confidence and that atmospheric neutrino tau appearance provides very strong constraints.

\section{Discussion}
\label{sec:discussion}
From figs.~\ref{fig:uti44_current}, \ref{fig:uti53_current} we see that currently with only the
information from the electron and muon row all values of $|U_{\tau i}|$ are allowed at $\Delta\chi^2<1$ in both scenarios up to large values of the matrix elements where strong constraints from the normalization of the tau row apply.
Even in the presence of a larger matrix the (3,3) matrix elements have a maximal value in order to satisfy the row normalization constraint. These maximal values of the matrix elements correspond to the case where all matrix elements involving the heavy mass states are zero (i.e.~$|U_{\tau 4}|=|U_{\tau5}|=0$).
As we add additional tau neutrino data sets, the constraints for large values of the matrix elements remain strong such that the upper limit on $|U_{\tau i}|$ remains similar, nearly independent of the data set added.
However not all data sets have a big impact.
From fig.~\ref{fig:uti44_current} we see that solar NC data does not add much constraining power to the information from the electron and muon row alone in both benchmark case considered. The sensitivity in both benchmark cases is similar, for clarity of the presentation we omit this line in fig.~\ref{fig:uti53_current}. The reason for this can be understood from the fact that the measurement of the total neutrino flux is more precise than the theoretical prediction (see appendix \ref{sec:exp} for more details) such that this data set does not add much information. Hence any future measurements of the solar neutrino flux will therefore need to be accompanied by equal improvements on the theoretical predictions to further constrain the tau row.
Similarly, experiments DONuT and FASERnu do not provide strong constraints in the (4,4) scenario.
In fact they provide only mild constraints ($\Delta \chi^2\lesssim 1$) for intermediate values of the matrix elements around $|U_{\tau i}|\sim 0.3-0.4$.
The reason for this is that in the (4,4) scenario the analytical oscillation probability is $P_{\text{CC scat}}^{(4,4)}=1-2 |U_{\tau4}|^2(|U_{\tau 1}|^2+|U_{\tau 2}|^2+|U_{\tau 3}|^2)$ which is then compared to the expected probability in the (3,3) case $P_{\text{CC scat}}^{(3,3)}=1$ (see appendix \ref{sec:exp} for more details).
Since large values of $|U_{\tau i}|,~i=1,2,3$ imply a small $|U_{\tau4}|$, as the tau row normalization needs to be fulfilled, the second part of the expression is small and $P_{\text{CC scat}}^{(4,4)}\sim 1$. On the other hand if the $|U_{\tau i}|,~i=1,2,3$ are small $|U_{\tau4}|$ needs to be big to fulfill the row normalization but again $P_{\text{CC scat}}^{(4,4)}\sim 1$. 
Only in the case of medium large matrix elements $|U_{\tau i}|\sim 0.3-0.4$ where all matrix elements are of similar order the second term of the oscillation probability is not anymore small but around 0.6. This leads to medium large values of all matrix elements to be slightly disfavored.
This cancellation is however not present in the (5,3) case due to the different expression of the oscillation probability (see appendix \ref{sec:exp}). In this case the CC scattering experiments constrain small values of the matrix elements and will in the future lead to competitive constraints.

OPERA is rather insensitive to medium values of the matrix elements in the (4,4) case and only strongly rules out small and large $|U_{\tau i}|$ whereas the exclusion power for small values is weaker in the (5,3) case. 
DUNE in the future will provide strong constraints. As DUNE will be sensitive not only to tau appearance in the beam but also inform the tau row from muon disappearance data it will improve the long baseline constraints from OPERA.
Astrophysical tau appearance is currently not very constraining due to the small data set of tau neutrino appearance ($\sim2$ events so far), however up-coming experiments will improve this data set and our forecasted results show that in both benchmark cases this channel is competitive with other data sets.
Atmospheric tau appearance currently excludes small matrix elements at the $\Delta \chi^2\gtrsim 9$ level in the (4,4) case and $\Delta \chi^2\gtrsim 2$ in the (5,3) case. In the future this data set will provide very strong constraints demonstrating that this channel is very useful to constrain unitarity.
In the (5,3) there are enough parameters to partially absorb the tau lepton effects in atmospheric appearance data, but in the (4,4) case this is no longer possible leading to stronger constraints in the (4,4) case.
CEvNS is currently not very constraining in both benchmark cases and only provides constraints around $\Delta \chi^2\sim 1$ level in the (4,4) case and below this level in the (5,3) such that we do not include this curve in our plot.
In the future this data set will be somewhat constraining in the (4,4) case and could disfavor small $|U_{\tau i}|$ at $\Delta \chi^2\gtrsim 4$.
In the (5,3) case CEvNS basically provides no significant constraint.
Atmospheric muon neutrino disappearance provides very strong constraints because of its high statistics. 
As this channel only constrains the $|U_{\tau 4}|$ matrix element no cancellations between different matrix elements are possible and we obtain a clean constraint on the deviation from the tau row normalization.
Even though not all categories of experiments analyzed have the same constraining power considering the global tau neutrino data set is important to obtain a complete picture of the tau sector.

In comparison to previous constraints on the tau row matrix elements which only included data from OPERA, atmospheric muon disappearance, and NOMAD data, we demonstrate that more tau data is currently available which improves the information on the tau row. 
It appeared that accelerator neutrinos DUNE might drive the tau row constraints in the future; we point, however, out that previously unaccounted for data from atmospheric tau appearance\footnote{Including, in principle, atmospheric tau neutrino appearance at DUNE \cite{Conrad:2010mh} which is not discussed in this article.} will be essential to obtain strong and complete constraints on the tau row together with more data from atmospheric muon disappearance. Also astrophysical tau appearance, CEvNS, as well as CC scattering experiments will further strengthen the constraints somewhat.
The dominant observables which will drive the sensitivity to the tau row matrix elements in the future are atmospheric muon disappearance and atmospheric tau appearance together with long baseline data from DUNE where muon disappearance as well as tau appearance data will be important.
In fact, we expect that in the future the tau row could be measured with comparable precision as the other rows with the inclusion of these data sets therefore improving the status of the tau row quite quickly.

Future constraints from individual probes such as long baseline accelerator, atmospheric disappearance, or atmospheric appearance, are comparable to the current combined constraint.
Thus the combined future constraint does not benefit substantially from the addition of data from scattering experiments or astrophysical $\nu_\tau$ appearance.
Finally, we also point out that at some point no improvement on the tau row can be achieved anymore without improving the other rows as well. This can be done with future experiments like JUNO, Hyper-Kamiokande, and DUNE which will in turn lead to progress on the unitarity constraints on the electron and muon row. 

Our results demonstrate the importance of new tau neutrino data sets which should be used to conduct a global UV fit also involving all available data sets for the electron, muon row which therefore allows to obtain a complete picture of the neutrino sector. Furthermore, UV in oscillations provides a complementary signature of heavy sterile neutrinos. 

To obtain a complete picture of the constraints on the unitarity of the leptonic mixing matrix a combination of all constraints from electroweak precision data, from direct searches, from oscillations needs to be conducted, however this goes beyond the scope of this manuscript.

\section{Conclusions}
\label{sec:conclusions}
The tau neutrino is the least well known particle of the SM. In fact, given previous studies of the data, large deviations from the SM expectation could be realized, making the tau sector an appealing window to new physics. In this manuscript we present new constraints on the elements of the tau row of the leptonic mixing matrix therefore probing a broad class of SM extensions which introduce unitarity violation and highlighting the relative importance of different experiments.

Phenomenologically motivated models of additional sterile neutrinos lead to apparent UV as the complete leptonic mixing matrix is unitary but the directly testable $3\times 3$ submatrix is not.
We studied two benchmark cases for UV parameterized by a $4\times 4$ unitary matrix with a sterile neutrino that is kinematically accessible but its oscillations are too fast to be resolved at detectors (i.e.~$m_4\in [10~\text{eV},~15~\text{MeV}]$), and a scenario where the sterile neutrino is not kinematically accessible ($m_4\gtrsim40$ MeV). 
For both benchmark scenarios we investigated for the first time several tau data sets in the literature which had been previously neglected including atmospheric tau appearance, astrophysical tau appearance, as well as CC and NC scattering experiments like CEvNS or with solar neutrinos. Combined with the data sets also explored by other unitarity studies like long baseline tau appearance data from OPERA and DUNE in the future, and atmospheric muon disappearance data we derive constraints on the individual tau matrix elements.
Therefore our results use all currently available tau data to showcase the relative constraining power of different experimental channels. 
We find that the introduction of the new data sets considerably improved the constraints on the tau row.

A precise measurement of the matrix elements can provide insights if the leptonic mixing matrix behaves similarly to the quark mixing matrix and if the $\mu-\tau$ symmetry is exactly realized in the mixing matrix which can guide flavor model building and constrain existing models \cite{Xing:2015fdg,Denton:2020exu}.

In summary, in this manuscript we have established the use of new tau neutrino data sets and have shown that the tau row is in a better shape than previously assumed in the literature.
We have also laid out the picture of tau neutrino unitarity constraints as they currently exist and will evolve over the coming decades.

\acknowledgments
We thank the anonymous referee for helpful comments.
We acknowledge support from the US Department of Energy under Grant Contract DE-SC0012704. We thank Enrique Fernandez-Martinez for pointing out the relevance of the atmospheric muon disappearance data in the (5,3) case.
Some of the figures and computations were done with \texttt{python} \cite{10.5555/1593511} and \texttt{matplotlib} \cite{Hunter:2007}.

\appendix
\section{Experiments}
\label{sec:exp}
In this appendix we describe the experimental data used in the analysis presented in the main text as well as the combinations of parameter constrained in the (4,4) and (5,3) cases.

When the experimental results are provided for both mass orderings we will assume for definiteness the normal ordering as current experiments provide a mild hint for the normal mass ordering in the three flavor scenario \cite{Esteban:2020cvm}.

\subsection{Atmospheric muon neutrino disappearance data}
Atmospheric muon neutrino disappearance data is sensitive to the mixing of heavy sterile neutrinos with tau neutrinos due to the presence of the SM matter effect \cite{Blennow:2018hto}. 

The best current constraints come from DeepCore \cite{Aartsen:2017bap} and Super-Kamiokande \cite{Abe:2014gda}
which constrain
\begin{align}
|U_{\tau4}|^2<0.15~\text{at 90$\%$ C.L.}~.
\end{align}

The constraint will be improved in the future with more data from these experiments as well as by KM3NeT/ORCA \cite{Aiello:2021aqp} and Hyper-Kamiokande \cite{Abe:2018uyc}. We assume these experiments find no evidence for unitary violation and use as future constraint
\begin{align}
|U_{\tau4}|^2< 0.048 ~\text{at 90$\%$ C.L.}\,,
\end{align}
which is achievable with, for example, with 5.6 Mton year exposure of Hyper-Kamiokande.
 
These constraints have been explicitly derived in the (4,4) scenarios and can be translated to the (5,3) scenario by replacing $|U_{\tau4}|^2\to |U_{\tau4}|^2+|U_{\tau5}|^2$ \cite{Fong:2016yyh,Blennow:2016jkn,Fong:2017gke}.
\subsection{Atmospheric tau appearance}
IceCube \cite{Aartsen:2019tjl} and Super-Kamiokande \cite{Li:2017dbe} have constrained the tau normalization $N_\tau$ defined as the ratio of the measured $\nu_\tau$ flux to the expected one assuming standard oscillations.
The results are 
\begin{align}
N_\tau&= 1.47\pm0.32 ~\text{from Super-Kamiokande},\\
N_\tau&=0.57^{+0.36}_{-0.34}~\text{from IceCube}.
\end{align}

As the signal events are rather centered in one energy and zenith angle bin we use a 1 bin analysis for true energies\footnote{We take the translation between reconstructed and true energy into account.}. 
We simulate the decrease of events in the energy-zenith angle space as two Gaussians for energy and zenith angle centered at 20 GeV (15 GeV), with $\sigma=20$ (10) GeV for IC (SK), and zenith angle centered at $\cos(\theta_z)=-1$ with $\sigma=0.3 (0.2)$ for IC (SK). We integrate over the full zenith angle range between -1 and 0 and between 5.6-56 GeV for IC and 3.50-70 GeV for SK which are the ranges these experiments are sensitive to tau appearance.
For these measurements the tau neutrinos are detected in CC processes coming from muon neutrinos produced in the atmosphere. The number of $\nu_\tau$ events in a bin is parameterized as 
\begin{align}
 n_{\nu_\tau}^{theo}=\sigma^{CC,SM}_\tau (UU^\dagger)_{\tau\tau} \phi_\mu^{SM} (UU^\dagger)_{\mu\mu} \tilde{P}_{\mu\tau} \frac{1}{(UU^\dagger)_{\mu\mu}(UU^\dagger)_{\tau\tau}}
\end{align}
where $\tilde{P}_{\mu\tau}$ is defined in eq.~\ref{eq:osciprob_tilde}.
The tau normalization can be written as the ratio of the measured tau neutrino flux to the theoretical expectation in a 3U framework,
\begin{align}
 N_\tau=\frac{\phi_{\nu_\tau}^{meas}}{\phi_{\nu_\tau}^{theo}}=\frac{\tilde{P}_{\mu\tau}^{UV}}{\tilde{P}_{\mu\tau}^{3U}}
\end{align}
where we use the best fit values from \cite{Esteban:2020cvm} in the calculation of $\tilde{P}_{\mu\tau}^{3U}$.
As it has been shown in \cite{Denton:2021rsa} the sensitivity to the tau normalization comes from SM matter effects which we parameterize with a constant density of $\rho=7 ~\text{g}/\text{cm}^3$. We then integrate over the energy and zenith angle to obtain $\tilde{P}_{\mu\tau}^{UV}$.

These constraints are going to be improved in the future by IceCube-Gen2, KM3NeT/ORCA \cite{Ishihara:2019aao,Eberl:2017plv,Aiello:2021jfn} and by Hyper-Kamiokande \cite{Abe:2018uyc}. For future measurements by IceCube-Gen2 and KM3NeT/ORCA we
assume that the tau normalization will be 
\begin{align}
 N_\tau=1\pm0.15 \text{ from KM3NeT/ORCA or IceCube-Gen2}
\end{align} which can be achieved for example with 1 year of KM3NeT/ORCA data \cite{Eberl:2017plv}. To simulate the improvement we use a 1-bin analysis with a Gaussian centered at 20 GeV, $\sigma=10$ GeV and $\cos\theta_z=-1$ $\sigma=0.3$ integrated over 6-70 GeV.
For the improvement by Hyper-Kamiokande we assume \cite{Abe:2018uyc}
\begin{align}
 N_\tau=1\pm 0.07 \text{ from Hyper-Kamiokande}
\end{align} and use the same Gaussians as for Super-Kamiokande.

\subsection{Astrophysical tau appearance}
Astrophysical neutrinos are produced in sources via the decays of pions, muons, or neutrons. These particles only produce electron and muon neutrinos such that the detection of astrophysical tau neutrinos indicates flavor change.

IceCube has detected 2 high-energy astrophysical tau neutrinos which leads to a flux (sum of neutrinos and anti-neutrinos) of
\cite{Abbasi:2020zmr}
\begin{align}
 \frac{d \phi_{\nu_\tau}}{dE}=3.0^{+2.2}_{-1.8}\left(\frac{E}{100\text{TeV}}\right)^{-2.87^{+0.21}_{-0.20}}\cdot 10^{-18}\text{ GeV}^{-1}\text{cm}^{-2}\text{s}^{-1}\text{sr}^{-1}
\end{align}
as well as a flux of astrophysical muon neutrinos which lead to tracks in the IceCube detector
(for the sum of neutrinos and anti-neutrinos) \cite{Stettner:2019tok}
\begin{align}
\frac{d \phi_{track}}{dE}=1.44^{+0.25}_{-0.24}\left(\frac{E}{100\text{TeV}}\right)^{-2.28^{+0.08}_{-0.09}}\cdot 10^{-18}\text{ GeV}^{-1}\text{cm}^{-2}\text{s}^{-1}\text{sr}^{-1}~.
\end{align}
The number of events of flavor $\alpha$ at IceCube is given by
\begin{align}
n_{\nu_\alpha}^{theo}=\sigma_\alpha^{CC,SM} (U U^\dagger)_{\alpha\alpha}\times \phi_p
(\xi (UU^\dagger)_{ee} P_{e\alpha} +(1-\xi) (UU^\dagger)_{\mu\mu}P_{\mu\alpha})
\end{align}
with the flux of the decaying particle $\phi_p$ ($p=$ pion, muon, or neutron). The factor $\xi$ parameterizes the 
 fraction of produced electron neutrinos per decaying particle from
$\xi=0$ (pion) to $\xi=1$ (neutron). Hence the product of $\phi_p \xi $ corresponds to the initial electron flux produced in the source (and $\phi_p (1-\xi) $ for the produced muon flux). As the origin of astrophysical neutrinos is unclear we will scan over
 $\xi$ between 0 and 1 in our analysis and not assume any prior on $\xi$.
Furthermore, we assume that no sterile neutrinos are produced at the source.

As neutrino oscillation lengths are much
smaller than astrophysical scales oscillations are
quickly averaged out and neutrinos have decohered when they arrive at the detector. Therefore the oscillation probability does not contain a kinematic factor and is
\begin{align}
P_{\alpha \beta}=\frac{1}{(U U^\dagger)_{\alpha\alpha}(UU^\dagger)_{\beta\beta}} \sum_{i=1}^{acc}|U_{\alpha i}|^2 |U_{\beta i}|^2~.
\end{align}
Furthermore, IceCube cannot easily distinguish between neutrinos and anti-neutrinos.
As we do not know $\phi_p$ we analyze the ratio of tau events to track event which is independent on $\phi_p$
\begin{equation}
\left( \frac{\phi_{\nu_\tau}}{\phi_{track}}\right)^{meas}=\frac{ \xi \sum_{i=1}^{acc}|U_{ei}|^2 |U_{\tau i}|^2+(1-\xi) \sum_{i=1}^{acc}|U_{\mu i}|^2 |U_{\tau i}|^2}{(1-\xi)\sum_{i=1}^{acc} |U_{\mu i}|^4+\xi\sum_{i=1}^{acc} |U_{ei}|^2|U_{\mu i}|^2 }~.
\end{equation}
This ratio is also independent of the CC cross section as 
for high energies this quantity is flavor independent \cite{Gandhi:1998ri}.
Finally,
we focus on the $E=100$ TeV bin of the measurement as it has the smallest uncertainty for all channels \cite{Stettner:2019tok}, this means that we do not take the uncertainty on the spectral index into account.

In the future also other neutrino telescopes can measure astrophysical $\nu_\tau$ appearance like IceCube-Gen2,
P-ONE, Baikal-GVD, ANITA/PUEO, GRAND, and POEMMA \cite{IceCube:2014gqr,Agostini:2020aar,Zaborov:2020idc,Allison:2020emr,Alvarez-Muniz:2018bhp,Krizmanic:2019hiq}.
We assume that with future observation of $\nu_\tau$ appearance the ratio will measured as
\begin{align}
\left( \frac{\phi_{\nu_\tau}}{\phi_{track}}\right)^{meas}=1\pm 0.14\,.
\end{align}

\subsection{Long baseline accelerator tau appearance}
Also human-made experiments with high energy neutrino beams like long baseline experiments are sensitive to tau neutrinos. 
The OPERA experiment directly detected the first tau neutrinos from oscillations in 2010 \cite{OPERA:2010pne} and DUNE will continue to contribute to the tau neutrino appearance data set in the future.

The OPERA experiment was located 730 km from the neutrino beam source which had an average neutrino energy of 17 GeV \cite{OPERA:2018nar}. 
We parameterize the number of observed events as
\begin{align}
 n_{\nu_{\tau}}^{meas}=A P(\nu_\mu\to\nu_\tau)+B\,,
\end{align}
where the constant factor $A$ includes the neutrino (anti-neutrino) flux times cross-section and $B$ is the background rate.
The collaboration provided the background events and the expected number of events for fiducial values of the oscillation parameters \cite{OPERA:2018nar} which allows to determine $A$.

The number of events depend on the matrix elements as
\begin{align}
 n_{\nu_{\tau}}^{theo}=\phi_\mu^{SM} (UU^\dagger)_{\mu\mu}
 \times \tilde{P}_{\mu\tau}\frac{1}{(UU^\dagger)_{\mu\mu}(UU^\dagger)_{\tau\tau}}\times\sigma_\tau^{CC,SM} (UU^\dagger)_{\tau\tau}
\end{align}
For these baselines and energies matter effects are important which we parameterize with a matter density of $\rho=3~\text{g}/\text{cm}^3$. In the (5,3) case we consider also the change of the Fermi constant as defined in eq.~\eqref{eq:gf}.
We numerically calculate the oscillation probability. 
Due to the low number of events we conduct a Poisson analysis of the number of events.
To validate our OPERA simulation we calculate the mass splitting in the (3,3) case $\Delta m_{32}^2= \left(2.7^{+0.9}_{-0.6}\right)\times 10^{-3}~\text{eV}^2$ which is in good agreement with the experimental result from \cite{OPERA:2018nar} $\Delta m_{32}^2=\left( 2.7^{+0.7}_{-0.6}\right)\times10^{-3}~\text{eV}^2$

In the future, DUNE will also provide valuable insights into the tau row. 
This experiment will also improve the other rows which by itself will advance the knowledge of the tau row using the unitarity conditions.
We use literature results for the constraints in the (4,4) and (5,3) case from \cite{deGouvea:2019ozk} (see also \cite{Ghoshal:2019pab}) which assume 3.5 years $\nu$ data+ 3.5 years $\bar{\nu}$ data.
For the (4,4) case using $\nu_e$-appearance, $\nu_\mu$-
disappearance data, and $\nu_\tau$-appearance \cite{deGouvea:2019ozk} 
the constraint is 
\begin{align}
 \sin^2\theta_{34}\leq 0.31 ~\text{at 95$\%$ C.L.\ .}
\end{align}
The constraint in the (5,3) case is \cite{deGouvea:2019ozk}
\begin{align}
 |U_{e3}|^2+|U_{\mu3}|^2+|U_{\tau3}|^2= 1^{+0.05}_{-0.06}~.
\end{align}
To avoid double counting we assume no improvements on the other rows when presenting future constraints on the tau row (see also discussion in the main text). 

Notice that DUNE will also have near detector from which the neutrino flux and cross section can be extracted. If these measured quantities are used in the analysis the cancellation of the prefactors in the number of expected events is not present. For example for tau neutrino appearance at the DUNE far detector the expected number of $\nu_\tau$ events is
\begin{align}
 n_{\nu_{\tau}}^{theo}=\phi_\mu^{meas} 
 \times \tilde{P}_{\mu\tau}\frac{1}{(UU^\dagger)_{\mu\mu}(UU^\dagger)_{\tau\tau}}\times\sigma_\tau^{CC,meas} ~.
\end{align}
In this case the matrix elements appear in several places, potentially increasing the sensitivity to them.

\subsection{Scattering experiments }
Scattering experiments which are not sensitive to oscillations still provide valuable constraints on the tau row as we demonstrate in the following. 
\subsubsection{Charged current scattering experiments}

Tau neutrino scattering experiments with a short baseline are sensitive to tau neutrino disappearance. If the $3\times3$ active-light mixing matrix is unitary the tau neutrino survival probability is 1 as for these experiments no SM oscillations have developed. 

\paragraph{DONuT}
The DONuT experiment discovered tau neutrinos for the first time in 2000 by using a high energy beam with mean energy $E\sim 100$ GeV and a detector close to the source with a baseline
$L\approx 40$ m \cite{Kodama:2007aa}.

We compare the number of observed events which is 9 to the number of predicted events from a Monte Carlo simulation which is 10. The difference is $-1\pm4$ \cite{Kodama:2007aa}. The number of expected events is parameterized as
\begin{align}
 n_{\nu_{\tau}}^{theo}=\phi_\tau^{SM} (UU^\dagger)_{\tau\tau}
 \times \tilde{P}_{\tau\tau}\frac{1}{(UU^\dagger)_{\tau\tau}(UU^\dagger)_{\tau\tau}}\times\sigma_\tau^{CC,SM} (UU^\dagger)_{\tau\tau}~.
\end{align}
In the (4,4) case the probability is 
\begin{align}
 \tilde{P}_{\tau\tau}=1-2|U_{\tau4}|^2(|U_{\tau1}|^2+|U_{\tau2}|^2+|U_{\tau3}|^2)~.
\end{align} 
For the (5,3) case the sensitivity to the matrix elements comes from the first term in oscillation probability
\begin{align}
 \tilde{P}_{\tau\tau}=||U_{\tau1}|^2+|U_{\tau2}|^2+|U_{\tau3}|^2|^2\neq 1~.
\end{align} 

\paragraph{FASERnu}
In the future FASERnu will use tau neutrinos from the LHC to measure tau cross sections\footnote{We focus on FASERnu but also other experiments at the LHC are sensitive to tau neutrinos like SBND$@$LHC \cite{SHiP:2020sos} however they expect less $\nu_\tau$ events; see \cite{Anchordoqui:2021ghd} for an overview of future forward physics facilities.}.
The baseline is $L=480$ m and like DONuT, FASERnu also has a broad neutrino energy spectrum but with a much larger mean energy $E\approx 1$ TeV . 
For FASERnu the number of expected detected events is $\sim11$ \cite{Abreu:2019yak} and we assume that the number of observed events equals the number of expected events.
The quantity to be constrained is the same as for DONuT in the previous subsection.

\paragraph{NOMAD}
NOMAD and CHORUS were experiments searching for tau appearance in the CERN wide-band $\nu_\mu$ beam with $E_\nu\sim 20-30$ GeV at a distance of 800 m. At these values of $L/E$ no SM oscillations have developed yet such that the results can be used to constrain sterile oscillations. The oscillation 
probabilities have been constrained by NOMAD to be\footnote{CHORUS has similar but slightly weaker constraints \cite{Eskut:2000de}.} \cite{Astier:2001yj}
\begin{align}
 & \tilde{P}(\nu_\mu\to \nu_\tau)<1.63\times10^{-4}~\rm{at }~ 90\%\text{C.L.}~,\\
 & \tilde{P}(\nu_e\to \nu_\tau)<0.74\times10^{-2}~\rm{at }~ 90\%\text{C.L.}~.
\end{align}
The analytical expression of the 
oscillation probability is
\begin{equation}
\tilde{P}(\nu_\mu\to \nu_\tau)= \left|\sum_{i=1}^{acc} U_{\mu i}^*U_{\tau i}\right|^2-2\Re\left(\sum_{j=heavy}^{acc}U_{\mu j}^* U_{\tau j}\sum_{i=1}^{3} U_{\mu i} U_{\tau i }^*\right)\,,
\label{eq:nomad prob}
\end{equation}
where the first sum goes over all kinematically accessible mass states, the second is over accessible mass states that are heavy compared to $m_1$, $m_2$, and $m_3$, and the third sum is over just the fist three mass states.
The pairing of the second two sums is to ensure that the relevant $\Delta m^2_{ji}L/4E$ is large enough to be oscillation averaged.
In the (4,4) case the first term (the so-called ``zero-distance term'') in eq.~\ref{eq:nomad prob} is just $\delta_{\mu\tau}$ while in the (5,3) case the second term (due to oscillation averaged) in eq.~\ref{eq:nomad prob} is zero.

In the (4,4) case these experiments provide a constraint on the second term in the oscillation probability which depends as $\tilde{P}_{\mu\tau}\propto U_{\mu4}U_{\tau4}$ (see also \cite{Coloma:2021uhq}).
With a prior on $U_{\mu4}$ from the normalization of the muon row from \cite{Hu:2020oba} which is compatible with 0 at $1\sigma$ this also means that the oscillation probability $P_{\mu\tau}$ is compatible with 0 at 1$\sigma$, hence we do not obtain a constraint on $U_{\tau4}$ from NOMAD alone (equal arguments apply to $P_{e\tau}$) in the (4,4) case. This statement can be easily understood as the effect of a sterile neutrino needs to be present in both the appearance and disappearance channels.

For the (5,3) case the second part of the probability is zero but the first part is $\leq 1$ as not all mass states are accessible.
The constraint in the (5,3) case only is on
\begin{align}
\tilde{P}_{\mu\tau}=|U_{\mu1}^*U_{\tau1}+U_{\mu2}^*U_{\tau2}+U_{\mu3}^*U_{\tau3}|^2=|U_{\mu4}^*U_{\tau4}+U_{\mu5}^*U_{\tau5}|^2\,.
\end{align}
by unitarity for the full matrix, and similarly for $P_{e\tau}$.
Then, for the same reason that we do not see sensitivity to the $U_{\tau i}$ elements given the constraints on the muon neutrino row, there is no sensitivity from the NOMAD data for the tau row in either the (4,4) or (5,3) cases.

\subsubsection{Neutral current scattering experiments}
Even though one does not identify tau neutrinos in neutral current measurements a non-unitary matrix still affects the measurement due to a reduction of the flux and detection cross section.
In fact, these measurements have potentially large statistics such that they could dominate the combined fits.
A crucial point about constraints of NC measurements is that the experimental result needs to be compared to a theory prediction of the number of events. This means the experiments benefit from theory predictions with small uncertainties.

\paragraph{CEvNS}
Coherent elastic neutrino nucleus scattering (CEvNS) describes a NC process where a neutrino scatters of a nucleus coherently \cite{Freedman:1973yd}. This process was first observed by the COHERENT collaboration in 2017 using low energy neutrinos from pion decays at rest at the SNS \cite{COHERENT:2017ipa}.
Although CEvNS detectors are not sensitive to SM oscillations and the initial flux does not contain $\nu_\tau$ their measurements still lead to constraints on the tau row as the detection process is NC.\footnote{See \cite{Miranda:2020syh} where UV constraints from CEvNS have been derived.}
This is an example of how flavor blind detection processes can constrain the tau row.
As an illustrative case we use the CEvNS measurement with the COHERENT CsI and Ar detectors\footnote{Notice that our results are independent of detector material as long as the same number of events with a similar uncertainty are detected.}. The initial flux contains muon and electron neutrinos from pion and muon decays at a stopped pion source.
Due to the finite lifetime and different energy distributions of the flux different initial flavors can be partially distinguished.
We conduct a simplified analysis using two timing bins integrated over the energy, the first bin ($t<1~\mu$s) contains events from the prompt pion decay which produces muon neutrinos, the second bin ($t\in (1,5)\mu$s) contains events from electron and muon neurinos from the delayed muon decay.

In our analysis we use the results from \cite{COHERENT:2020iec, Zettlemoyer:2020kgh,COHERENT:2020ybo} for the COHERENT Ar detector and \cite{M7slides} for the newest results from the COHERENT CsI detector. These references provide the number of expected events, measured events and background events.
Due to the short baselines of $\mathcal{O}(20~\text{m})$ and low energies $E\sim 30$ MeV at these detectors no oscillations have developed in the standard picture and the number of coming from flavor $\alpha~(\alpha=e,\mu)$ are
\begin{equation}
n_{NC}^{CEvNS}=\sigma^{NC,SM}\sum_{\alpha=e}^\mu\phi_\alpha^{\rm SM}\sum_{\beta=e}^\tau\left[\left|\sum_{i=1}^{acc} U_{\alpha i}^*U_{\beta i}\right|^2-2\Re\left(\sum_{j=heavy}^{acc}U_{\alpha j}^* U_{\beta j}\sum_{i=1}^{3} U_{\alpha i} U_{\beta i }^*\right)\right](UU^\dagger)_{\beta\beta}\,,
\end{equation}
where the square bracket term is the same as the probability for NOMAD in eq.~\ref{eq:nomad prob}, but the event rate now has an additional factor of $(UU^\dagger)_{\beta\beta}$ since this is a NC interaction.

We conduct a simple $\chi^2$ analysis\footnote{It has been shown in \cite{Denton:2020hop} that a careful statistical treatment is necessary to obtain robust constraints on new physics parameters. This conclusion was derived for the example of non standard interactions. For UV one would need to conduct a detailed study to see if these conclusions hold. We leave this for future work.} comparing the observed number of events to the predicted number of events
including constant nuisance parameters on the signal normalization, the steady state background normalization, and the normalization of the beam related neutrons with uncertainties provided in \cite{COHERENT:2020iec, Zettlemoyer:2020kgh,COHERENT:2020ybo,M7slides}.

In the future we forecast that events corresponding to 5 time the exposure of the CsI detector are detected. We assume an improvement of the signal normalization from $13\%$ to $7\%$, possible for example with better quenching factor and efficiency measurements, and steady state and neutron background normalizations of $1\%$.
Currently the signal normalization uncertainty is driven by the uncertainty on the neutrino flux. In the future this uncertainty can be reduced with the
D$_2$O detector \cite{Akimov:2021nkt} installed at the SNS by measuring the $\nu_e+d$ CC interactions. This measurement will however be impacted by the presence of UV as well which leads to a change in the CC cross section. Hence to make use of this data in a UV framework with future data, this effect needs to be included as well. 

\paragraph{NC measurement of solar neutrinos}
The NC measurement of solar neutrinos by SNO proved to be crucial for establishing neutrino oscillations. Even though this measurement was insensitive to the neutrino flavors some of the detected neutrinos are tau neutrinos as all mass eigenstates have a sizeable $\nu_\tau$ component. By comparing the measured NC flux to the theoretical prediction we can hence obtain constraints on tau neutrinos.\footnote{Previous UV fits also took solar data into account however as a ratio of the CC flux over the NC flux.
However, since we are focusing on the tau row, the only information that would come to $U_{\tau2}$ is from a knowledge of $U_{e2}$ and $U_{\mu2}$, but $U_{e2}$ cannot be separated from $U_{e1}$ based on KamLand data; data from JUNO, however, will be able to break the $\nu_1\Longleftrightarrow\nu_2$ degeneracy separate from solar data.}

 From \cite{Aharmim:2011vm} using the standard solar model the theoretical expectation for the $^8$Be flux is 
\begin{align}
 \phi_\nu^{\rm{theo}}&= (5.88\pm0.65)\times10^6\ \rm{cm}^{-2}\rm{s}^{-1}~\rm{( BPS09(GS))}, ~\\
 \phi_\nu^{\rm{theo}}&= (4.85\pm0.58)\times10^6\ \rm{cm}^{-2}\rm{s}^{-1}~\rm{(BPS09(AGSS09))}
\end{align}
which needs to be compared to the measurement by SNO \cite{Aharmim:2011vm}
\begin{equation}
 \phi_\nu^{\rm{exp}}=(5.25\pm0.16(\rm{stat.})^{+0.11}_{-0.13}(\rm{syst.}))\times10^6\ \rm{cm}^{-2}\rm{s}^{-1}~.
\end{equation}
As the two theoretical values slightly disagree we use their average value with an uncertainty which also includes the $1\sigma$ values of each of the individual predictions
\begin{equation}
 \phi_\nu^{\rm{pred}}=(5.365\pm1.1) \times10^6\ \rm{cm}^{-2}\rm{s}^{-1}~.
\end{equation}

The number of events is 
\begin{align}
 n_{NC}^{theo} =\phi^{SM} (U U^\dagger)_{ee}\times \sum_{j=1}^{acc}\sum_{i=1}^{acc} \sigma^{NC, SM} |( U^\dagger U)_{ij}|^2 \times \tilde{P}_{ei}\times \frac{1}{(UU^\dagger)_{ee}}
\end{align}
Note that the sums in the $U^\dagger U$ go over $\alpha\in\{e,\mu,\tau\}$ in the term from the cross section in the numerator. Since the states are properly normalized $(U^\dagger U)_{ii}=1$ if summed over $\alpha\in\{e,\mu,\tau\,s\}$.
At $E\gtrsim10$ MeV, $\tilde{P}_{e1}\simeq0$, $\tilde{P}_{e2}\simeq|U_{e1}|^2+|U_{e2}|^2$, $\tilde{P}_{e3}\simeq|U_{e3}|^2$.
In the (4,4) case additionally 
$\tilde{P}_{e4}=|U_{e4}|^2$. Notice that since the sterile mass splitting is large enough to not be relevant for the transitions the Sun this probability is the same as in vacuum.
In our analysis we are comparing the measured flux to the theory prediction, hence
\begin{align}
 \phi^{meas}_{NC}=\phi^{SM}_{NC} \times \sum_{i=1}^{acc}\sum_{j=1}^{acc} |( U^\dagger U)_{ij}|^2 \times \tilde{P}_{ei}~.
\end{align}

\begin{table}
\centering
\caption{Overview of the experiments considered in this analysis including the way we analyze their data, the processes related to production and detection of neutrinos and if there is a cancellation of the normalization terms in the number of events. For forecasted DUNE events, which rely on theoretical input for the cross section and flux from simulations, there are cancellations in the denominator, however as soon as near detector data is used in the analysis there will not be a cancellation anymore as measured quantities will get compared (see main text for more details).}
\begin{tabular}{c||c|c|c|c}
experiment&measurement&production&detection&cancellation\\\hline
atmospheric $\nu_{\mu}$ disappearance&$n_{\nu_\mu}^{meas}$ vs $n_{\nu_\mu}^{theo}$&CC&CC&yes\\
atmospheric $\nu_\tau$ appearance& $n_{\nu_\tau}^{meas}$ vs $n_{\nu_\tau}^{theo} $&CC&CC&yes\\
astrophysical $\nu_\tau$ appearance& $n_{\nu_\tau}^{meas}$ vs $n_{\nu_\mu}^{meas}$&CC&CC&yes \\
solar data&$\phi_{NC}^{meas}$ vs $\phi_{NC}^{theo}$ &CC&NC&no \\
DONuT/FASERnu&$n_{\nu_\tau}^{meas}$ vs $n_{\nu_\tau}^{theo} $&CC&CC&yes\\
LBL $\nu_\tau$ appearance (OPERA)&$n_{\nu_\tau}^{meas}$ vs $n_{\nu_\tau}^{theo} $&CC&CC&yes\\
LBL $\nu_\tau$ appearance (DUNE)& $n_{\nu_\tau}^{meas}$ vs $n_{\nu_\tau}^{theo} $&CC&CC&yes ($\rightarrow$ no)\\
LBL $\nu_\mu$ disappearance (DUNE)& $n_{\nu_\mu}^{meas}$ vs $n_{\nu_\mu}^{theo} $&CC&CC&yes ($\rightarrow$ no)\\
CEvNS& $n_{NC}^{meas}$ vs $n_{NC}^{theo}$&CC&NC&no
\end{tabular}
\label{tab:cancellation}
\end{table}

As the theoretical uncertainty on the flux is larger than the experimental one the constraints from solar data are rather weak, see sec.~\ref{sec:results}. This means in order to obtain strong constraints from future measurements of the solar neutrino flux, for example by JUNO \cite{An:2015jdp}, an improvement of the theoretical prediction of the flux is necessary.

It should be noted that elastic scattering processes are also sensitive to tau matrix elements however the flavor blind cross section is smaller than one involving electron neutrinos. For this reason we focus on the NC measurement in our analysis.

Another important difference arises for NC processes, namely the approximate translation between the constraints in the (4,4) and (5,3) scenario which is for CC processes exact up to $\mathcal{O}(\theta_s)^4$ \cite{Fong:2016yyh,Blennow:2016jkn,Fong:2017gke} is for NC processes only valid up to $\mathcal{O}(\theta_s)^2$.

\subsection{Summary of experiments}
\label{sec:sum_exp}
We provide in table \ref{tab:cancellation} a summary of all (current and future) experiments considered in the main text.

\bibliographystyle{JHEP}
\bibliography{main}

\providecommand{\href}[2]{#2}\begingroup\raggedright\begin{thebibliography}{10}

\bibitem{DONUT:2000fbd}
{\scshape DONUT} collaboration, \emph{{Observation of tau neutrino
  interactions}},
  \href{https://doi.org/10.1016/S0370-2693(01)00307-0}{\emph{Phys. Lett. B}
  {\bfseries 504} (2001) 218}
  [\href{https://arxiv.org/abs/hep-ex/0012035}{{\ttfamily hep-ex/0012035}}].

\bibitem{ATLAS:2012yve}
{\scshape ATLAS} collaboration, \emph{{Observation of a new particle in the
  search for the Standard Model Higgs boson with the ATLAS detector at the
  LHC}}, \href{https://doi.org/10.1016/j.physletb.2012.08.020}{\emph{Phys.
  Lett. B} {\bfseries 716} (2012) 1}
  [\href{https://arxiv.org/abs/1207.7214}{{\ttfamily 1207.7214}}].

\bibitem{CMS:2012qbp}
{\scshape CMS} collaboration, \emph{{Observation of a New Boson at a Mass of
  125 GeV with the CMS Experiment at the LHC}},
  \href{https://doi.org/10.1016/j.physletb.2012.08.021}{\emph{Phys. Lett. B}
  {\bfseries 716} (2012) 30} [\href{https://arxiv.org/abs/1207.7235}{{\ttfamily
  1207.7235}}].

\bibitem{Pontecorvo:1957qd}
B.~Pontecorvo, \emph{{Inverse beta processes and nonconservation of lepton
  charge}}, {\emph{Zh. Eksp. Teor. Fiz.} {\bfseries 34} (1957) 247}.

\bibitem{Maki:1962mu}
Z.~Maki, M.~Nakagawa and S.~Sakata, \emph{{Remarks on the unified model of
  elementary particles}}, \href{https://doi.org/10.1143/PTP.28.870}{\emph{Prog.
  Theor. Phys.} {\bfseries 28} (1962) 870}.

\bibitem{Parke:2015goa}
S.~Parke and M.~Ross-Lonergan, \emph{{Unitarity and the three flavor neutrino
  mixing matrix}},
  \href{https://doi.org/10.1103/PhysRevD.93.113009}{\emph{Phys. Rev. D}
  {\bfseries 93} (2016) 113009}
  [\href{https://arxiv.org/abs/1508.05095}{{\ttfamily 1508.05095}}].

\bibitem{Ellis:2020ehi}
S.A.R.~Ellis, K.J.~Kelly and S.W.~Li, \emph{{Leptonic Unitarity Triangles}},
  \href{https://doi.org/10.1103/PhysRevD.102.115027}{\emph{Phys. Rev. D}
  {\bfseries 102} (2020) 115027}
  [\href{https://arxiv.org/abs/2004.13719}{{\ttfamily 2004.13719}}].

\bibitem{Ellis:2020hus}
S.A.R.~Ellis, K.J.~Kelly and S.W.~Li, \emph{{Current and Future Neutrino
  Oscillation Constraints on Leptonic Unitarity}},
  \href{https://doi.org/10.1007/JHEP12(2020)068}{\emph{JHEP} {\bfseries 12}
  (2020) 068} [\href{https://arxiv.org/abs/2008.01088}{{\ttfamily
  2008.01088}}].

\bibitem{Hu:2020oba}
Z.~Hu, J.~Ling, J.~Tang and T.~Wang, \emph{{Global oscillation data analysis on
  the $3\nu$ mixing without unitarity}},
  \href{https://doi.org/10.1007/JHEP01(2021)124}{\emph{JHEP} {\bfseries 01}
  (2021) 124} [\href{https://arxiv.org/abs/2008.09730}{{\ttfamily
  2008.09730}}].

\bibitem{Cabibbo:1963yz}
N.~Cabibbo, \emph{{Unitary Symmetry and Leptonic Decays}},
  \href{https://doi.org/10.1103/PhysRevLett.10.531}{\emph{Phys. Rev. Lett.}
  {\bfseries 10} (1963) 531}.

\bibitem{Kobayashi:1973fv}
M.~Kobayashi and T.~Maskawa, \emph{{CP Violation in the Renormalizable Theory
  of Weak Interaction}}, \href{https://doi.org/10.1143/PTP.49.652}{\emph{Prog.
  Theor. Phys.} {\bfseries 49} (1973) 652}.

\bibitem{Charles:2015gya}
J.~Charles et~al., \emph{{Current status of the Standard Model CKM fit and
  constraints on $\Delta F=2$ New Physics}},
  \href{https://doi.org/10.1103/PhysRevD.91.073007}{\emph{Phys. Rev. D}
  {\bfseries 91} (2015) 073007}
  [\href{https://arxiv.org/abs/1501.05013}{{\ttfamily 1501.05013}}].

\bibitem{ArkaniHamed:1998vp}
N.~Arkani-Hamed, S.~Dimopoulos, G.R.~Dvali and J.~March-Russell,
  \emph{{Neutrino masses from large extra dimensions}},
  \href{https://doi.org/10.1103/PhysRevD.65.024032}{\emph{Phys. Rev. D}
  {\bfseries 65} (2001) 024032}
  [\href{https://arxiv.org/abs/hep-ph/9811448}{{\ttfamily hep-ph/9811448}}].

\bibitem{ArkaniHamed:1998sj}
N.~Arkani-Hamed and S.~Dimopoulos, \emph{{New origin for approximate symmetries
  from distant breaking in extra dimensions}},
  \href{https://doi.org/10.1103/PhysRevD.65.052003}{\emph{Phys. Rev. D}
  {\bfseries 65} (2002) 052003}
  [\href{https://arxiv.org/abs/hep-ph/9811353}{{\ttfamily hep-ph/9811353}}].

\bibitem{Bhattacharya:2009nu}
S.~Bhattacharya, P.~Dey and B.~Mukhopadhyaya, \emph{{Unitarity violation in
  sequential neutrino mixing in a model of extra dimensions}},
  \href{https://doi.org/10.1103/PhysRevD.80.075013}{\emph{Phys. Rev. D}
  {\bfseries 80} (2009) 075013}
  [\href{https://arxiv.org/abs/0907.0099}{{\ttfamily 0907.0099}}].

\bibitem{Minkowski:1977sc}
P.~Minkowski, \emph{{$\mu \to e\gamma$ at a Rate of One Out of $10^{9}$ Muon
  Decays?}}, \href{https://doi.org/10.1016/0370-2693(77)90435-X}{\emph{Phys.
  Lett. B} {\bfseries 67} (1977) 421}.

\bibitem{Schechter:1980gr}
J.~Schechter and J.W.F.~Valle, \emph{{Neutrino Masses in SU(2) x U(1)
  Theories}}, \href{https://doi.org/10.1103/PhysRevD.22.2227}{\emph{Phys. Rev.
  D} {\bfseries 22} (1980) 2227}.

\bibitem{Foot:1988aq}
R.~Foot, H.~Lew, X.G.~He and G.C.~Joshi, \emph{{Seesaw Neutrino Masses Induced
  by a Triplet of Leptons}}, \href{https://doi.org/10.1007/BF01415558}{\emph{Z.
  Phys. C} {\bfseries 44} (1989) 441}.

\bibitem{Bolton:2019pcu}
P.D.~Bolton, F.F.~Deppisch and P.S.~Bhupal~Dev, \emph{{Neutrinoless double beta
  decay versus other probes of heavy sterile neutrinos}},
  \href{https://doi.org/10.1007/JHEP03(2020)170}{\emph{JHEP} {\bfseries 03}
  (2020) 170} [\href{https://arxiv.org/abs/1912.03058}{{\ttfamily
  1912.03058}}].

\bibitem{Qian:2013ora}
X.~Qian, C.~Zhang, M.~Diwan and P.~Vogel, \emph{{Unitarity Tests of the
  Neutrino Mixing Matrix}},  \href{https://arxiv.org/abs/1308.5700}{{\ttfamily
  1308.5700}}.

\bibitem{Forero:2021azc}
D.V.~Forero, C.~Giunti, C.A.~Ternes and M.~Tortola, \emph{{Nonunitary neutrino
  mixing in short and long-baseline experiments}},
  \href{https://doi.org/10.1103/PhysRevD.104.075030}{\emph{Phys. Rev. D}
  {\bfseries 104} (2021) 075030}
  [\href{https://arxiv.org/abs/2103.01998}{{\ttfamily 2103.01998}}].

\bibitem{Antusch:2006vwa}
S.~Antusch, C.~Biggio, E.~Fernandez-Martinez, M.B.~Gavela and J.~Lopez-Pavon,
  \emph{{Unitarity of the Leptonic Mixing Matrix}},
  \href{https://doi.org/10.1088/1126-6708/2006/10/084}{\emph{JHEP} {\bfseries
  10} (2006) 084} [\href{https://arxiv.org/abs/hep-ph/0607020}{{\ttfamily
  hep-ph/0607020}}].

\bibitem{Fernandez-Martinez:2015hxa}
E.~Fernandez-Martinez, J.~Hernandez-Garcia, J.~Lopez-Pavon and M.~Lucente,
  \emph{{Loop level constraints on Seesaw neutrino mixing}},
  \href{https://doi.org/10.1007/JHEP10(2015)130}{\emph{JHEP} {\bfseries 10}
  (2015) 130} [\href{https://arxiv.org/abs/1508.03051}{{\ttfamily
  1508.03051}}].

\bibitem{Fernandez-Martinez:2016lgt}
E.~Fernandez-Martinez, J.~Hernandez-Garcia and J.~Lopez-Pavon, \emph{{Global
  constraints on heavy neutrino mixing}},
  \href{https://doi.org/10.1007/JHEP08(2016)033}{\emph{JHEP} {\bfseries 08}
  (2016) 033} [\href{https://arxiv.org/abs/1605.08774}{{\ttfamily
  1605.08774}}].

\bibitem{Mention:2011rk}
G.~Mention, M.~Fechner, T.~Lasserre, T.A.~Mueller, D.~Lhuillier, M.~Cribier
  et~al., \emph{{The Reactor Antineutrino Anomaly}},
  \href{https://doi.org/10.1103/PhysRevD.83.073006}{\emph{Phys. Rev. D}
  {\bfseries 83} (2011) 073006}
  [\href{https://arxiv.org/abs/1101.2755}{{\ttfamily 1101.2755}}].

\bibitem{Giunti:2010zu}
C.~Giunti and M.~Laveder, \emph{{Statistical Significance of the Gallium
  Anomaly}}, \href{https://doi.org/10.1103/PhysRevC.83.065504}{\emph{Phys. Rev.
  C} {\bfseries 83} (2011) 065504}
  [\href{https://arxiv.org/abs/1006.3244}{{\ttfamily 1006.3244}}].

\bibitem{Kostensalo:2019vmv}
J.~Kostensalo, J.~Suhonen, C.~Giunti and P.C.~Srivastava, \emph{{The gallium
  anomaly revisited}},
  \href{https://doi.org/10.1016/j.physletb.2019.06.057}{\emph{Phys. Lett. B}
  {\bfseries 795} (2019) 542}
  [\href{https://arxiv.org/abs/1906.10980}{{\ttfamily 1906.10980}}].

\bibitem{LSND:2001aii}
{\scshape LSND} collaboration, \emph{{Evidence for neutrino oscillations from
  the observation of $\bar{\nu}_e$ appearance in a $\bar{\nu}_\mu$ beam}},
  \href{https://doi.org/10.1103/PhysRevD.64.112007}{\emph{Phys. Rev. D}
  {\bfseries 64} (2001) 112007}
  [\href{https://arxiv.org/abs/hep-ex/0104049}{{\ttfamily hep-ex/0104049}}].

\bibitem{MiniBooNE:2018esg}
{\scshape MiniBooNE} collaboration, \emph{{Significant Excess of ElectronLike
  Events in the MiniBooNE Short-Baseline Neutrino Experiment}},
  \href{https://doi.org/10.1103/PhysRevLett.121.221801}{\emph{Phys. Rev. Lett.}
  {\bfseries 121} (2018) 221801}
  [\href{https://arxiv.org/abs/1805.12028}{{\ttfamily 1805.12028}}].

\bibitem{IceCube:2020tka}
{\scshape IceCube} collaboration, \emph{{Searching for eV-scale sterile
  neutrinos with eight years of atmospheric neutrinos at the IceCube Neutrino
  Telescope}}, \href{https://doi.org/10.1103/PhysRevD.102.052009}{\emph{Phys.
  Rev. D} {\bfseries 102} (2020) 052009}
  [\href{https://arxiv.org/abs/2005.12943}{{\ttfamily 2005.12943}}].

\bibitem{Fukugita:1986hr}
M.~Fukugita and T.~Yanagida, \emph{{Baryogenesis Without Grand Unification}},
  \href{https://doi.org/10.1016/0370-2693(86)91126-3}{\emph{Phys. Lett. B}
  {\bfseries 174} (1986) 45}.

\bibitem{Blennow:2016jkn}
M.~Blennow, P.~Coloma, E.~Fernandez-Martinez, J.~Hernandez-Garcia and
  J.~Lopez-Pavon, \emph{{Non-Unitarity, sterile neutrinos, and Non-Standard
  neutrino Interactions}},
  \href{https://doi.org/10.1007/JHEP04(2017)153}{\emph{JHEP} {\bfseries 04}
  (2017) 153} [\href{https://arxiv.org/abs/1609.08637}{{\ttfamily
  1609.08637}}].

\bibitem{Fong:2016yyh}
C.S.~Fong, H.~Minakata and H.~Nunokawa, \emph{{A framework for testing leptonic
  unitarity by neutrino oscillation experiments}},
  \href{https://doi.org/10.1007/JHEP02(2017)114}{\emph{JHEP} {\bfseries 02}
  (2017) 114} [\href{https://arxiv.org/abs/1609.08623}{{\ttfamily
  1609.08623}}].

\bibitem{Fong:2017gke}
C.S.~Fong, H.~Minakata and H.~Nunokawa, \emph{{Non-unitary evolution of
  neutrinos in matter and the leptonic unitarity test}},
  \href{https://doi.org/10.1007/JHEP02(2019)015}{\emph{JHEP} {\bfseries 02}
  (2019) 015} [\href{https://arxiv.org/abs/1712.02798}{{\ttfamily
  1712.02798}}].

\bibitem{Esteban:2020cvm}
I.~Esteban, M.C.~Gonzalez-Garcia, M.~Maltoni, T.~Schwetz and A.~Zhou,
  \emph{{The fate of hints: updated global analysis of three-flavor neutrino
  oscillations}}, \href{https://doi.org/10.1007/JHEP09(2020)178}{\emph{JHEP}
  {\bfseries 09} (2020) 178}
  [\href{https://arxiv.org/abs/2007.14792}{{\ttfamily 2007.14792}}].

\bibitem{Planck:2018vyg}
{\scshape Planck} collaboration, \emph{{Planck 2018 results. VI. Cosmological
  parameters}},
  \href{https://doi.org/10.1051/0004-6361/201833910}{\emph{Astron. Astrophys.}
  {\bfseries 641} (2020) A6}
  [\href{https://arxiv.org/abs/1807.06209}{{\ttfamily 1807.06209}}].

\bibitem{Ballett:2019bgd}
P.~Ballett, T.~Boschi and S.~Pascoli, \emph{{Heavy Neutral Leptons from
  low-scale seesaws at the DUNE Near Detector}},
  \href{https://doi.org/10.1007/JHEP03(2020)111}{\emph{JHEP} {\bfseries 03}
  (2020) 111} [\href{https://arxiv.org/abs/1905.00284}{{\ttfamily
  1905.00284}}].

\bibitem{Mikheyev:1985zog}
S.P.~Mikheyev and A.Y.~Smirnov, \emph{{Resonance Amplification of Oscillations
  in Matter and Spectroscopy of Solar Neutrinos}}, {\emph{Sov. J. Nucl. Phys.}
  {\bfseries 42} (1985) 913}.

\bibitem{Denton:2021rsa}
P.B.~Denton, \emph{{Tau neutrino identification in atmospheric neutrino
  oscillations without particle identification or unitarity}},
  \href{https://doi.org/10.1103/PhysRevD.104.113003}{\emph{Phys. Rev. D}
  {\bfseries 104} (2021) 113003}
  [\href{https://arxiv.org/abs/2109.14576}{{\ttfamily 2109.14576}}].

\bibitem{Fernandez-Martinez:2007iaa}
E.~Fernandez-Martinez, M.B.~Gavela, J.~Lopez-Pavon and O.~Yasuda,
  \emph{{CP-violation from non-unitary leptonic mixing}},
  \href{https://doi.org/10.1016/j.physletb.2007.03.069}{\emph{Phys. Lett. B}
  {\bfseries 649} (2007) 427}
  [\href{https://arxiv.org/abs/hep-ph/0703098}{{\ttfamily hep-ph/0703098}}].

\bibitem{Blennow:2018hto}
M.~Blennow, E.~Fernandez-Martinez, J.~Gehrlein, J.~Hernandez-Garcia and
  J.~Salvado, \emph{{IceCube bounds on sterile neutrinos above 10 eV}},
  \href{https://doi.org/10.1140/epjc/s10052-018-6282-2}{\emph{Eur. Phys. J. C}
  {\bfseries 78} (2018) 807}
  [\href{https://arxiv.org/abs/1803.02362}{{\ttfamily 1803.02362}}].

\bibitem{Aartsen:2017bap}
{\scshape IceCube} collaboration, \emph{{Search for sterile neutrino mixing
  using three years of IceCube DeepCore data}},
  \href{https://doi.org/10.1103/PhysRevD.95.112002}{\emph{Phys. Rev. D}
  {\bfseries 95} (2017) 112002}
  [\href{https://arxiv.org/abs/1702.05160}{{\ttfamily 1702.05160}}].

\bibitem{Abe:2014gda}
{\scshape Super-Kamiokande} collaboration, \emph{{Limits on sterile neutrino
  mixing using atmospheric neutrinos in Super-Kamiokande}},
  \href{https://doi.org/10.1103/PhysRevD.91.052019}{\emph{Phys. Rev. D}
  {\bfseries 91} (2015) 052019}
  [\href{https://arxiv.org/abs/1410.2008}{{\ttfamily 1410.2008}}].

\bibitem{Aiello:2021aqp}
S.~Aiello et~al., \emph{{Sensitivity to light sterile neutrino mixing
  parameters with KM3NeT/ORCA}},
  \href{https://arxiv.org/abs/2107.00344}{{\ttfamily 2107.00344}}.

\bibitem{MINOS:2017cae}
{\scshape MINOS+} collaboration, \emph{{Search for sterile neutrinos in MINOS
  and MINOS+ using a two-detector fit}},
  \href{https://doi.org/10.1103/PhysRevLett.122.091803}{\emph{Phys. Rev. Lett.}
  {\bfseries 122} (2019) 091803}
  [\href{https://arxiv.org/abs/1710.06488}{{\ttfamily 1710.06488}}].

\bibitem{NOvA:2017geg}
{\scshape NOvA} collaboration, \emph{{Search for active-sterile neutrino mixing
  using neutral-current interactions in NOvA}},
  \href{https://doi.org/10.1103/PhysRevD.96.072006}{\emph{Phys. Rev. D}
  {\bfseries 96} (2017) 072006}
  [\href{https://arxiv.org/abs/1706.04592}{{\ttfamily 1706.04592}}].

\bibitem{NOvA:2021smv}
{\scshape NOvA} collaboration, \emph{{Search for active-sterile antineutrino
  mixing using neutral-current interactions with the NOvA experiment}},
  \href{https://arxiv.org/abs/2106.04673}{{\ttfamily 2106.04673}}.

\bibitem{T2K:2019efw}
{\scshape T2K} collaboration, \emph{{Search for light sterile neutrinos with
  the T2K far detector Super-Kamiokande at a baseline of 295 km}},
  \href{https://doi.org/10.1103/PhysRevD.99.071103}{\emph{Phys. Rev. D}
  {\bfseries 99} (2019) 071103}
  [\href{https://arxiv.org/abs/1902.06529}{{\ttfamily 1902.06529}}].

\bibitem{Aartsen:2019tjl}
{\scshape IceCube} collaboration, \emph{{Measurement of Atmospheric Tau
  Neutrino Appearance with IceCube DeepCore}},
  \href{https://doi.org/10.1103/PhysRevD.99.032007}{\emph{Phys. Rev. D}
  {\bfseries 99} (2019) 032007}
  [\href{https://arxiv.org/abs/1901.05366}{{\ttfamily 1901.05366}}].

\bibitem{Li:2017dbe}
{\scshape Super-Kamiokande} collaboration, \emph{{Measurement of the tau
  neutrino cross section in atmospheric neutrino oscillations with
  Super-Kamiokande}},
  \href{https://doi.org/10.1103/PhysRevD.98.052006}{\emph{Phys. Rev. D}
  {\bfseries 98} (2018) 052006}
  [\href{https://arxiv.org/abs/1711.09436}{{\ttfamily 1711.09436}}].

\bibitem{Ishihara:2019aao}
{\scshape IceCube} collaboration, \emph{{The IceCube Upgrade -- Design and
  Science Goals}}, \href{https://doi.org/10.22323/1.358.1031}{\emph{PoS}
  {\bfseries ICRC2019} (2020) 1031}
  [\href{https://arxiv.org/abs/1908.09441}{{\ttfamily 1908.09441}}].

\bibitem{Eberl:2017plv}
{\scshape KM3NeT} collaboration, \emph{{Tau neutrino appearance with
  KM3NeT/ORCA}}, \href{https://doi.org/10.22323/1.301.1025}{\emph{PoS}
  {\bfseries ICRC2017} (2018) 1025}.

\bibitem{Aiello:2021jfn}
{\scshape KM3NeT} collaboration, \emph{{Determining the Neutrino Mass Ordering
  and Oscillation Parameters with KM3NeT/ORCA}},
  \href{https://arxiv.org/abs/2103.09885}{{\ttfamily 2103.09885}}.

\bibitem{Abe:2018uyc}
{\scshape Hyper-Kamiokande} collaboration, \emph{{Hyper-Kamiokande Design
  Report}},  \href{https://arxiv.org/abs/1805.04163}{{\ttfamily 1805.04163}}.

\bibitem{Razzaque:2009kq}
S.~Razzaque and A.Y.~Smirnov, \emph{{Flavor conversion of cosmic neutrinos from
  hidden jets}}, \href{https://doi.org/10.1007/JHEP03(2010)031}{\emph{JHEP}
  {\bfseries 03} (2010) 031} [\href{https://arxiv.org/abs/0912.4028}{{\ttfamily
  0912.4028}}].

\bibitem{Palladino:2018qgi}
A.~Palladino, C.~Mascaretti and F.~Vissani, \emph{{The importance of observing
  astrophysical tau neutrinos}},
  \href{https://doi.org/10.1088/1475-7516/2018/08/004}{\emph{JCAP} {\bfseries
  08} (2018) 004} [\href{https://arxiv.org/abs/1804.04965}{{\ttfamily
  1804.04965}}].

\bibitem{Abbasi:2020zmr}
{\scshape IceCube} collaboration, \emph{{Measurement of Astrophysical Tau
  Neutrinos in IceCube's High-Energy Starting Events}},
  \href{https://arxiv.org/abs/2011.03561}{{\ttfamily 2011.03561}}.

\bibitem{Stettner:2019tok}
{\scshape IceCube} collaboration, \emph{{Measurement of the Diffuse
  Astrophysical Muon-Neutrino Spectrum with Ten Years of IceCube Data}},
  \href{https://doi.org/10.22323/1.358.1017}{\emph{PoS} {\bfseries ICRC2019}
  (2020) 1017} [\href{https://arxiv.org/abs/1908.09551}{{\ttfamily
  1908.09551}}].

\bibitem{IceCube:2014gqr}
{\scshape IceCube} collaboration, \emph{{IceCube-Gen2: A Vision for the Future
  of Neutrino Astronomy in Antarctica}},
  \href{https://arxiv.org/abs/1412.5106}{{\ttfamily 1412.5106}}.

\bibitem{OPERA:2018nar}
{\scshape OPERA} collaboration, \emph{{Final Results of the OPERA Experiment on
  $\nu_\tau$ Appearance in the CNGS Neutrino Beam}},
  \href{https://doi.org/10.1103/PhysRevLett.120.211801}{\emph{Phys. Rev. Lett.}
  {\bfseries 120} (2018) 211801}
  [\href{https://arxiv.org/abs/1804.04912}{{\ttfamily 1804.04912}}].

\bibitem{deGouvea:2019ozk}
A.~De~Gouv\^ea, K.J.~Kelly, G.V.~Stenico and P.~Pasquini, \emph{{Physics with
  Beam Tau-Neutrino Appearance at DUNE}},
  \href{https://doi.org/10.1103/PhysRevD.100.016004}{\emph{Phys. Rev. D}
  {\bfseries 100} (2019) 016004}
  [\href{https://arxiv.org/abs/1904.07265}{{\ttfamily 1904.07265}}].

\bibitem{Astier:2001yj}
{\scshape NOMAD} collaboration, \emph{{Final NOMAD results on muon-neutrino
  ---\ensuremath{>} tau-neutrino and electron-neutrino ---\ensuremath{>}
  tau-neutrino oscillations including a new search for tau-neutrino appearance
  using hadronic tau decays}},
  \href{https://doi.org/10.1016/S0550-3213(01)00339-X}{\emph{Nucl. Phys. B}
  {\bfseries 611} (2001) 3}
  [\href{https://arxiv.org/abs/hep-ex/0106102}{{\ttfamily hep-ex/0106102}}].

\bibitem{Eskut:2000de}
{\scshape CHORUS} collaboration, \emph{{New results from a search for nu/mu
  --\ensuremath{>} nu/tau and nu/e --\ensuremath{>} nu/tau oscillation}},
  \href{https://doi.org/10.1016/S0370-2693(00)01317-4}{\emph{Phys. Lett. B}
  {\bfseries 497} (2001) 8}.

\bibitem{Kodama:2007aa}
{\scshape DONuT} collaboration, \emph{{Final tau-neutrino results from the
  DONuT experiment}},
  \href{https://doi.org/10.1103/PhysRevD.78.052002}{\emph{Phys. Rev. D}
  {\bfseries 78} (2008) 052002}
  [\href{https://arxiv.org/abs/0711.0728}{{\ttfamily 0711.0728}}].

\bibitem{Abreu:2019yak}
{\scshape FASER} collaboration, \emph{{Detecting and Studying High-Energy
  Collider Neutrinos with FASER at the LHC}},
  \href{https://doi.org/10.1140/epjc/s10052-020-7631-5}{\emph{Eur. Phys. J. C}
  {\bfseries 80} (2020) 61} [\href{https://arxiv.org/abs/1908.02310}{{\ttfamily
  1908.02310}}].

\bibitem{Anchordoqui:2021ghd}
L.A.~Anchordoqui et~al., \emph{{The Forward Physics Facility: Sites,
  Experiments, and Physics Potential}},
  \href{https://arxiv.org/abs/2109.10905}{{\ttfamily 2109.10905}}.

\bibitem{NOvA:2021nfi}
{\scshape NOvA, R. Group} collaboration, \emph{{An Improved Measurement of
  Neutrino Oscillation Parameters by the NOvA Experiment}},
  \href{https://arxiv.org/abs/2108.08219}{{\ttfamily 2108.08219}}.

\bibitem{T2K:2021xwb}
{\scshape T2K} collaboration, \emph{{Improved constraints on neutrino mixing
  from the T2K experiment with $\mathbf{3.13\times10^{21}}$ protons on
  target}}, \href{https://doi.org/10.1103/PhysRevD.103.112008}{\emph{Phys. Rev.
  D} {\bfseries 103} (2021) 112008}
  [\href{https://arxiv.org/abs/2101.03779}{{\ttfamily 2101.03779}}].

\bibitem{Denton:2020uda}
P.B.~Denton, J.~Gehrlein and R.~Pestes, \emph{{$CP$ -Violating Neutrino
  Nonstandard Interactions in Long-Baseline-Accelerator Data}},
  \href{https://doi.org/10.1103/PhysRevLett.126.051801}{\emph{Phys. Rev. Lett.}
  {\bfseries 126} (2021) 051801}
  [\href{https://arxiv.org/abs/2008.01110}{{\ttfamily 2008.01110}}].

\bibitem{Chatterjee:2020kkm}
S.S.~Chatterjee and A.~Palazzo, \emph{{Nonstandard Neutrino Interactions as a
  Solution to the $NO\nu A$ and T2K Discrepancy}},
  \href{https://doi.org/10.1103/PhysRevLett.126.051802}{\emph{Phys. Rev. Lett.}
  {\bfseries 126} (2021) 051802}
  [\href{https://arxiv.org/abs/2008.04161}{{\ttfamily 2008.04161}}].

\bibitem{Conrad:2010mh}
J.~Conrad, A.~de~Gouvea, S.~Shalgar and J.~Spitz, \emph{{Atmospheric Tau
  Neutrinos in a Multi-kiloton Liquid Argon Detector}},
  \href{https://doi.org/10.1103/PhysRevD.82.093012}{\emph{Phys. Rev. D}
  {\bfseries 82} (2010) 093012}
  [\href{https://arxiv.org/abs/1008.2984}{{\ttfamily 1008.2984}}].

\bibitem{Xing:2015fdg}
Z.-z.~Xing and Z.-h.~Zhao, \emph{{A review of
  \ensuremath{\mu}-\ensuremath{\tau} flavor symmetry in neutrino physics}},
  \href{https://doi.org/10.1088/0034-4885/79/7/076201}{\emph{Rept. Prog. Phys.}
  {\bfseries 79} (2016) 076201}
  [\href{https://arxiv.org/abs/1512.04207}{{\ttfamily 1512.04207}}].

\bibitem{Denton:2020exu}
P.B.~Denton, \emph{{A Return To Neutrino Normalcy}},
  \href{https://arxiv.org/abs/2003.04319}{{\ttfamily 2003.04319}}.

\bibitem{10.5555/1593511}
G.~Van~Rossum and F.L.~Drake, \emph{Python 3 Reference Manual}, CreateSpace,
  Scotts Valley, CA (2009).

\bibitem{Hunter:2007}
J.D.~Hunter, \emph{Matplotlib: A 2d graphics environment},
  \href{https://doi.org/10.1109/MCSE.2007.55}{\emph{Computing in Science \&
  Engineering} {\bfseries 9} (2007) 90}.

\bibitem{Gandhi:1998ri}
R.~Gandhi, C.~Quigg, M.H.~Reno and I.~Sarcevic, \emph{{Neutrino interactions at
  ultrahigh-energies}},
  \href{https://doi.org/10.1103/PhysRevD.58.093009}{\emph{Phys. Rev. D}
  {\bfseries 58} (1998) 093009}
  [\href{https://arxiv.org/abs/hep-ph/9807264}{{\ttfamily hep-ph/9807264}}].

\bibitem{Agostini:2020aar}
{\scshape P-ONE} collaboration, \emph{{The Pacific Ocean Neutrino Experiment}},
  \href{https://doi.org/10.1038/s41550-020-1182-4}{\emph{Nature Astron.}
  {\bfseries 4} (2020) 913} [\href{https://arxiv.org/abs/2005.09493}{{\ttfamily
  2005.09493}}].

\bibitem{Zaborov:2020idc}
{\scshape Baikal-GVD} collaboration, \emph{{High-energy neutrino astronomy and
  the Baikal-GVD neutrino telescope}},  in \emph{{5th International Conference
  on Particle Physics and Astrophysics}}, 11, 2020
  [\href{https://arxiv.org/abs/2011.09209}{{\ttfamily 2011.09209}}].

\bibitem{Allison:2020emr}
Q.~Abarr et~al., \emph{{The Payload for Ultrahigh Energy Observations (PUEO): A
  White Paper}},  \href{https://arxiv.org/abs/2010.02892}{{\ttfamily
  2010.02892}}.

\bibitem{Alvarez-Muniz:2018bhp}
{\scshape GRAND} collaboration, \emph{{The Giant Radio Array for Neutrino
  Detection (GRAND): Science and Design}},
  \href{https://doi.org/10.1007/s11433-018-9385-7}{\emph{Sci. China Phys. Mech.
  Astron.} {\bfseries 63} (2020) 219501}
  [\href{https://arxiv.org/abs/1810.09994}{{\ttfamily 1810.09994}}].

\bibitem{Krizmanic:2019hiq}
{\scshape POEMMA} collaboration, \emph{{POEMMA: Probe Of Extreme
  Multi-Messenger Astrophysics}},
  \href{https://doi.org/10.1051/epjconf/201921006008}{\emph{EPJ Web Conf.}
  {\bfseries 210} (2019) 06008}.

\bibitem{OPERA:2010pne}
{\scshape OPERA} collaboration, \emph{{Observation of a first $\nu_\tau$
  candidate in the OPERA experiment in the CNGS beam}},
  \href{https://doi.org/10.1016/j.physletb.2010.06.022}{\emph{Phys. Lett. B}
  {\bfseries 691} (2010) 138}
  [\href{https://arxiv.org/abs/1006.1623}{{\ttfamily 1006.1623}}].

\bibitem{Ghoshal:2019pab}
A.~Ghoshal, A.~Giarnetti and D.~Meloni, \emph{{On the role of the $\nu_{\tau}$
  appearance in DUNE in constraining standard neutrino physics and beyond}},
  \href{https://doi.org/10.1007/JHEP12(2019)126}{\emph{JHEP} {\bfseries 12}
  (2019) 126} [\href{https://arxiv.org/abs/1906.06212}{{\ttfamily
  1906.06212}}].

\bibitem{SHiP:2020sos}
{\scshape SHiP} collaboration, \emph{{SND@LHC}},
  \href{https://arxiv.org/abs/2002.08722}{{\ttfamily 2002.08722}}.

\bibitem{Coloma:2021uhq}
P.~Coloma, J.~L\'opez-Pav\'on, S.~Rosauro-Alcaraz and S.~Urrea, \emph{{New
  physics from oscillations at the DUNE near detector, and the role of
  systematic uncertainties}},
  \href{https://arxiv.org/abs/2105.11466}{{\ttfamily 2105.11466}}.

\bibitem{Freedman:1973yd}
D.Z.~Freedman, \emph{{Coherent Neutrino Nucleus Scattering as a Probe of the
  Weak Neutral Current}},
  \href{https://doi.org/10.1103/PhysRevD.9.1389}{\emph{Phys. Rev. D} {\bfseries
  9} (1974) 1389}.

\bibitem{COHERENT:2017ipa}
{\scshape COHERENT} collaboration, \emph{{Observation of Coherent Elastic
  Neutrino-Nucleus Scattering}},
  \href{https://doi.org/10.1126/science.aao0990}{\emph{Science} {\bfseries 357}
  (2017) 1123} [\href{https://arxiv.org/abs/1708.01294}{{\ttfamily
  1708.01294}}].

\bibitem{Miranda:2020syh}
O.G.~Miranda, D.K.~Papoulias, O.~Sanders, M.~T\'ortola and J.W.F.~Valle,
  \emph{{Future CEvNS experiments as probes of lepton unitarity and
  light-sterile neutrinos}},
  \href{https://doi.org/10.1103/PhysRevD.102.113014}{\emph{Phys. Rev. D}
  {\bfseries 102} (2020) 113014}
  [\href{https://arxiv.org/abs/2008.02759}{{\ttfamily 2008.02759}}].

\bibitem{COHERENT:2020iec}
{\scshape COHERENT} collaboration, \emph{{First Measurement of Coherent Elastic
  Neutrino-Nucleus Scattering on Argon}},
  \href{https://doi.org/10.1103/PhysRevLett.126.012002}{\emph{Phys. Rev. Lett.}
  {\bfseries 126} (2021) 012002}
  [\href{https://arxiv.org/abs/2003.10630}{{\ttfamily 2003.10630}}].

\bibitem{Zettlemoyer:2020kgh}
J.C.~Zettlemoyer, \emph{{First Detection of Coherent Elastic Neutrino-nucleus
  Scattering on an Argon Target}}, Ph.D. thesis, Indiana U., Bloomington
  (main), Indiana U., Bloomington (main), 5, 2020.
\newblock 10.5967/3wza-6w73.

\bibitem{COHERENT:2020ybo}
{\scshape COHERENT} collaboration, \emph{{COHERENT Collaboration data release
  from the first detection of coherent elastic neutrino-nucleus scattering on
  argon}},  \href{https://arxiv.org/abs/2006.12659}{{\ttfamily 2006.12659}}.

\bibitem{M7slides}
D.~Pershey, ``{New Results from the COHERENT CsI[Na] Detector}.''
  \href{https://indico.cern.ch/event/943069/contributions/4066386/attachments/2143826/3613080/20201116_M7s_CsI.pdf}{presented
  at Magnificent CEvNS 2020 }, 2020.

\bibitem{Denton:2020hop}
P.B.~Denton and J.~Gehrlein, \emph{{A Statistical Analysis of the COHERENT Data
  and Applications to New Physics}},
  \href{https://doi.org/10.1007/JHEP04(2021)266}{\emph{JHEP} {\bfseries 04}
  (2021) 266} [\href{https://arxiv.org/abs/2008.06062}{{\ttfamily
  2008.06062}}].

\bibitem{Akimov:2021nkt}
{\scshape COHERENT} collaboration, \emph{{A D$_{2}$O detector for flux
  normalization of a pion decay-at-rest neutrino source}},
  \href{https://arxiv.org/abs/2104.09605}{{\ttfamily 2104.09605}}.

\bibitem{Aharmim:2011vm}
{\scshape SNO} collaboration, \emph{{Combined Analysis of all Three Phases of
  Solar Neutrino Data from the Sudbury Neutrino Observatory}},
  \href{https://doi.org/10.1103/PhysRevC.88.025501}{\emph{Phys. Rev. C}
  {\bfseries 88} (2013) 025501}
  [\href{https://arxiv.org/abs/1109.0763}{{\ttfamily 1109.0763}}].

\bibitem{An:2015jdp}
{\scshape JUNO} collaboration, \emph{{Neutrino Physics with JUNO}},
  \href{https://doi.org/10.1088/0954-3899/43/3/030401}{\emph{J. Phys. G}
  {\bfseries 43} (2016) 030401}
  [\href{https://arxiv.org/abs/1507.05613}{{\ttfamily 1507.05613}}].

\end{thebibliography}\endgroup

\end{document}